\documentclass{aa}  
\usepackage{aas_macros}
\usepackage{xcolor}

\usepackage{graphicx}
\usepackage{txfonts}
\usepackage{lipsum}
\usepackage{subcaption}         
\usepackage{lscape}             
\usepackage{placeins}                                           
\usepackage{hyperref}
\newcommand*{\M}{\ensuremath{\text{M}}}
\renewcommand*{\P}{\ensuremath{\text{P}}}

\newcommand*{\RM}{R_\text{M}}

\begin{document}
   \title{Tidal disruptions of rubble piles: The case of Phobos}

   \author{H. Agrusa\inst{1}
        \and P. Michel\inst{1,2}
        }

   \institute{Universit\'e C\^ote d'Azur, Observatoire de la C\^ote d'Azur, CNRS, Laboratoire Lagrange, Nice, France\\
             \email{hagrusa@oca.eu}
            \and The University of Tokyo, Department of Systems Innovation, School of Engineering, Tokyo 113-0033, Japan\\ }

   \date{Received 5 November, 2025 / Accepted 9 January 2026}
 
  \abstract
   {Many small satellites in the Solar System have sub-synchronous orbits, meaning their orbits are decaying due to tidal dissipation. Unless they have substantial material strength, they will eventually tidally disrupt before reaching their planet's surface.}
   {We studied the fate of rubble-pile satellites as they migrate inward, with a particular focus on the case of Phobos.} 
   {We used a combination of analytic estimates and numerical simulations to determine the failure mode and tidal disruption distance of a Phobos-like satellite, as a function of its shape and cohesive strength.}
   {Both analytically and numerically, we identify a regime for satellites with low cohesive strengths whereby their surface can be tidally stripped without undergoing internal failure. Our numerical simulations demonstrate that Phobos will be destroyed beyond 2 Mars radii if it has a bulk strength similar to those estimated for small bodies recently visited by spacecraft.}
   {Based on our results and some additional arguments, we suggest that previous studies on the fate of Phobos have overestimated its strength, and therefore underestimated its tidal disruption distance. We also speculate that if Phobos undergoes some tidal stripping, its ultimate fate may be determined by runaway collisional erosion rather than a pure tidal disruption. However, the ultimate tidal disruption distance for Phobos will depend on its unknown internal structure and bulk material properties, which will be constrained by JAXA's Martian Moons eXploration (MMX) mission and its IDEFIX rover. These results have implications for theories about the origin and evolution of the Martian moons and for tidal disruptions of other small, irregularly shaped satellites.}


   \maketitle
\section{Introduction}
\nolinenumbers
Phobos orbits within the synchronous radius, meaning its orbital period is shorter than Mars' rotation period, causing its orbit to decay due to the Martian tides. When Phobos reaches its Roche limit, tidal stresses are expected to disrupt the body and form a ring \citep{Black2015}. Today, there are already visible signs of Phobos' eventual tidal demise on its surface. The enigmatic grooves on Phobos' surface may be caused by tidal stresses leading to surface fractures \citep{Hurford2016, ChengBin2022}. There is evidence of mass wasting due to Phobos orbital decay \citep{Shi2016}. In addition, granular creep motion driven by time-varying tidal and rotational forces may explain the origin of the blue units on Phobos' surface \citep{Ballouz2019}. The precise distance at which Phobos will undergo tidal disruption is highly dependent on Phobos' shape, which is well known, and Phobos' material properties and internal structure, which are poorly constrained \citep{Holsapple2006,Holsapple2008,Black2015}.

The origin of Mars' two moons remains an open question. Their shapes, spectra, and surfaces resemble asteroids, suggesting that they could be captured bodies \citep[e.g.,][]{Burns1978}. However, their circular and coplanar orbits challenge this hypothesis because two random captures are unlikely to result in such a configuration \citep{Burns1992} and tides cannot circularize their orbits within the Solar System's lifetime \citep{Rosenblatt2011}. Alternatively, the two moons would naturally have circular and coplanar orbits if they accreted from a giant impact-generated debris disk \citep[e.g.,][]{Craddock2011}. The Japanese Martian Moons eXploration (MMX) mission will launch in 2026 to explore Phobos in great detail, including returning a sample to Earth in 2031, and also conduct several close flybys of Deimos \citep{Kuramoto2022}. A principal goal of the mission is to determine the origin of the Martian Moons. In addition to constraining Phobos origin, MMX will also greatly improve our understanding of Phobos' fate by constraining its internal structure and material properties through gravity science, shape modeling, spectral analysis, and in situ investigations with the IDEFIX rover \citep[e.g.,][]{Matsumoto2021,Nakamura2021,Barucci2025,Michel2022b,Ulamec2025}. 

Here, we explore the fate of a rubble-pile Phobos as it migrates inward as a function of its shape and material properties. In Sect. \ref{sec:analytic} we estimate Phobos' Roche limit analytically and also demonstrate that material may be stripped from its surface well outside of the orbital distance at which it will undergo internal failure. Then, in Sect. \ref{sec:numerical} we numerically model the inward migration and tidal disruption of Phobos and verify the analytic predictions. The results are discussed and contrasted with previous studies in Sect. \ref{sec:discussion}, followed by conclusions in Sect. \ref{sec:conclusions}.

\section{Analytic estimate for the Roche limit}\label{sec:analytic}
Here, we provide two independent analytic estimates for the Roche limit for a uniform-density, ellipsoidal satellite. The first is a simple estimate of the orbital distance at which the tidal and centrifugal accelerations overcome self-gravity, which defines the point at which loose material would be stripped from the surface. The second is an analysis of the point at which the internal stresses overcome the Drucker-Prager failure criterion, which considers the role of frictional and cohesive forces and was first derived by \cite{Holsapple2008}.

\subsection{The rigid-body Roche limit}\label{subsec:RRL}
The classic rigid-body Roche limit is defined as the orbital distance, $d$, at which the tidal acceleration is equal to self-gravity at the sub-primary (or anti-primary) point on the surface of a spherical satellite \citep[e.g.,][]{Aggarwal1974}. For a spherical satellite that has a mass, radius, and density of $M_2,R_2,\rho_2$ orbiting a primary with $M_1,R_1,\rho_1$, this can be written as
 
\begin{align}
    a_\text{tides} &= a_\text{grav} \\
    \frac{M_1}{(d-R_2)^2} - \frac{M_1}{d^2} &= \frac{M_2}{R_2^2}. \\ \label{eq:classic_roche}
\end{align}
If the radius of the satellite is small compared to the Roche distance (i.e., $R_2 \ll d$), then this expression can be simplified with a first-order Taylor expansion and we recover the classic rigid-body Roche limit, which we denote $d_\text{RRL}$,
\begin{align}
    \frac{M_1}{d^2}(1-\frac{R_2}{d})^{-2} - \frac{M_1}{d^2} &= \frac{M_2}{R_2^2} \\
    \frac{M_1}{d^2}(1+2\frac{R_2}{d}) - \frac{M_1}{d^2} &= \frac{M_2}{R_2^2} \\
    2\frac{M_1R_2}{d^3} &= \frac{M_2}{R_2^2} \\
    d^3 &= 2\frac{M_1}{M_2}R_2^3\\
    d_\text{RRL} &= \bigg(2\frac{\rho_1}{\rho_2}\bigg)^{1/3}R_1.\\
\end{align}
For equal-density spheres, $d_\text{RRL}{\sim}1.259R_1$. In the case of Phobos orbiting Mars, this corresponds to $d_\text{RRL}{\sim}1.6R_\text{M}$, where $R_\text{M}$ is the radius of Mars. This value for $d_\text{RRL}$ is often used as a nominal disruption radius for studies related to the formation of a ring following the tidal disruption of Phobos \citep{Hesselbrock2017,Madeira2023b}. However, this estimate neglects the centrifugal acceleration of a synchronously rotating Phobos. In addition, the nonspherical shape of Phobos (or any irregularly shaped satellite, generally speaking) will change the self-gravitational acceleration and strongly influence the location of the rigid-body Roche limit. 

To approximate this effect, we introduce the “second-order rigid-body Roche limit,” which we refer to as $d_\text{2oRRL}$. It is similar to the classic rigid-body Roche limit ($d_\text{RRL}$), but includes the centrifugal acceleration of a synchronous satellite along with a second-order approximation for self-gravity. In this case, the secondary is described as a uniform-density ellipsoid, having a mass, $M_2$, bulk density, $\rho_2$, moments of inertia, $A\leq B \leq C$, corresponding to semiaxis lengths $a \geq b \geq c$, and a volume-equivalent radius of $R_2=(a b c)^{1/3}$. We use MacCullagh's formula to approximate the satellite's self-gravity to second order at the sub-primary point, assuming that the satellite is tidally locked,
\begin{equation}
    a_\text{grav} = -\frac{GM_2}{a^2} - \frac{3G}{2a^4}\big(B+C-2A\big),
\end{equation}
where $G$ is the gravitational constant. The tidal acceleration is the relative gravitational acceleration due to the primary between the sub-primary point and the satellite's center,
\begin{equation}
    a_\text{tides} = \frac{GM_1}{(d-a)^2} -\frac{GM_1}{d^2},
\end{equation}
where $d$ is the distance between the primary and satellite. Finally, for a tidally locked satellite, the centrifugal acceleration can be related to the mean motion, $n$:
\begin{equation}
    a_\text{rot} = n^2a = \frac{G(M_1+M_2)}{d^3}a
.\end{equation}
The Roche limit is then the location where these three accelerations sum to zero:
\begin{align}
a_\text{grav}+a_\text{tides}+a_\text{rot} &= 0 \\
-\frac{GM_2}{a^2} - \frac{3G}{2a^4}\big(B+C-2A\big) + \frac{GM_1}{(d-a)^2} -\frac{GM_1}{d^2} + \frac{G(M_1+M_2)}{d^3}a &= 0 \\
.\end{align}
In general, this equation can be solved numerically for $d$, as a function of the primary and secondary properties. In the case of Phobos, we can assume that $a\ll d$, which allows us to simplify the $GM_1/(d-a)^2$ term with a first-order Taylor expansion. We can also assume that $M_1+M_2\approx M_1$ and rewrite the moments of inertia of the satellite (A,B,C) in terms of its dynamically equivalent equal-volume ellipsoid (DEEVE) semiaxis lengths (a,b,c). With these simplifications we arrive at
\begin{equation}
    -\frac{M_2}{a^2}-\frac{3M_2(2a^2-b^2-c^2)}{10a^4} + \frac{3M_1a}{d^3} = 0.
\end{equation}
Solving for $d$ we have
\begin{align}
    d^3 &= -\frac{30M_1 a^5}{M_2(-16a^2+3b^2+3c^2)} \\
    d &= \bigg(\frac{-30 a^4}{bc(-16a^2+3b^2+3c^2)}\bigg)^{1/3} \bigg(\frac{\rho_1}{\rho_2}\bigg)^{1/3}R_1.\\
\end{align}
Then, we can introduce the aspect ratios $\alpha=c/a$ and $\beta=b/a$, and with some simplifications we arrive at a simple expression for $d_\text{2oRRL}$:

\begin{equation}
    \label{eq:d_shed}
    d_\text{2oRRL} = \bigg(\frac{10}{\alpha\beta(16-3\alpha^2-3\beta^2)}\bigg)^{1/3} \bigg(\frac{3\rho_1}{\rho_2}\bigg)^{1/3}R_1.
\end{equation}

\begin{figure}
    \centering
    \includegraphics[width=\linewidth]{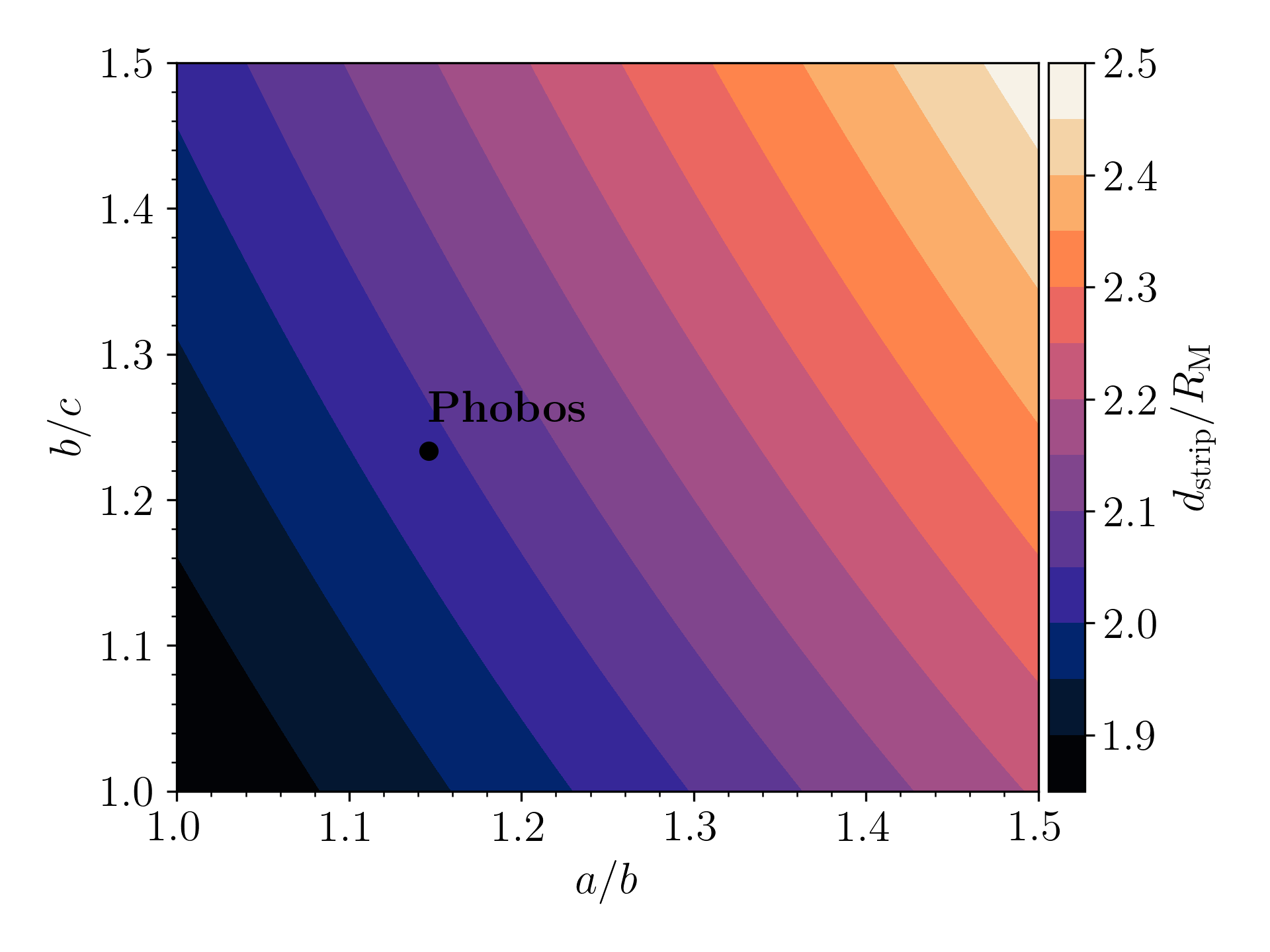}
    \caption{\label{fig:phobos_RRL} Second-order rigid-body Roche limit, referred to as $d_\text{strip}$, for a uniform-density triaxial ellipsoid around Mars, as a function of its axis ratios $a/b$ and $b/c$. With Phobos' present-day bulk density and shape, the surface acceleration at the sub-Mars point will become zero at ${\sim}2.03R_\text{M}$, meaning that material could be freely stripped from the surface.}
\end{figure}

At this distance, the net acceleration at the sub-primary point is zero, meaning material could be freely stripped from the surface so we interchangeably refer to $d_\text{2oRRL}$ as $d_\text{strip}$. We can see in the case of a spherical satellite that the first term in parenthesis simplifies to 1 and we recover the classic rigid-body Roche limit (including rotation) of $(3\rho_1/\rho_2)^{1/3}R_1$. For the case of Phobos around Mars, with a bulk density for Phobos of $1.861 \text{g cm}^{-3}$ and semiaxis lengths of $(a,b,c)=(12.95,11.30,9.16) \text{ km}$ \citep{Ernst2023a}, this results in $d_\text{strip}\approx2.03R_\text{M}$. In Fig.\ \ref{fig:phobos_RRL} we show $d_\text{strip}$ as a function of the ellipsoidal shape of a body that has a Phobos density around Mars. 

\subsection{The role of cohesion and friction}\label{subsec:druckerprager}
The previous derivation neglects the material properties of the satellite; namely, the body's friction angle and cohesive forces, which can strengthen it against tidal disruption. The Drucker-Prager yield criterion, which is commonly used in the study of granular materials, can be written as
\begin{equation}
    \label{eq:drucker}
     \frac{1}{6}\bigg((\sigma_x-\sigma_y)^{2} + (\sigma_y-\sigma_z)^{2} + (\sigma_z-\sigma_x)^{2})\bigg) \leq \bigg(k - s(\sigma_x+\sigma_y+\sigma_z)\bigg)^2,
 \end{equation} 
 where {the left side is the second invariant of the deviator stresses, commonly referred to as $J_2$}, $\sigma_x, \sigma_y,\sigma_z$ correspond to the body's three principal stresses, and $k$ and $s$ are constants related to the cohesion and friction of the body. In terms of the cohesive strength, $c$, and the angle of internal friction, $\phi$, $k$ and $s$ are defined as

\begin{align}
k &= \frac{6c\cos\phi}{\sqrt{3}(3-\sin\phi)}, \\
s &= \frac{2\sin\phi}{\sqrt{3}(3-\sin\phi)}.
\end{align}

For a given value of $k$ and $s$, an approximate solution can be derived using the volume-averaged stress over the body. From Eq. 13 of \cite{Holsapple2008}, the volume-averaged normal stresses on a rotating ellipsoidal body, with self-gravity and tidal forces can be written as
\begin{align}
\overline{\sigma}_x &= \bigg[\rho_2\omega^2 - 2\pi\rho_2^2GA_x + \frac{8\pi}{3}G\rho_1\rho_2\bigg(\frac{d}{R_1}\bigg)^{-3}\bigg]\frac{a^2}{5} \\
\overline{\sigma}_y &= \bigg[\rho_2\omega^2 - 2\pi\rho_2^2GA_y - \frac{4\pi}{3}G\rho_1\rho_2\bigg(\frac{d}{R_1}\bigg)^{-3}\bigg]\frac{b^2}{5} \\
\overline{\sigma}_z &= \bigg[-2\pi\rho_2^2GA_z - \frac{4\pi}{3}G\rho_1\rho_2\bigg(\frac{d}{R_1}\bigg)^{-3}\bigg]\frac{c^2}{5}.
\end{align}

Here, the terms $A_x$, $A_y$, and $A_z$ are nondimensional terms that only depend on the shape of the body. They are written as
\begin{align}
A_x &= \alpha\beta\int_0^\infty\frac{\mathrm{d}u}{(u+1)^{3/2}(u+\beta^2)^{1/2}(u+\alpha^2)^{1/2}}, \\
A_y &= \alpha\beta\int_0^\infty\frac{\mathrm{d}u}{(u+1)^{1/2}(u+\beta^2)^{3/2}(u+\alpha^2)^{1/2}}, \\
A_z &= \alpha\beta\int_0^\infty\frac{\mathrm{d}u}{(u+1)^{1/2}(u+\beta^2)^{1/2}(u+\alpha^2)^{3/2}}.
\end{align}

For a synchronous satellite, we have $\omega^2=n^2=G(M_1+M_2)/d^3\approx GM_1/d^3=4\pi G\rho_1 R_1^3/3d^3$, which allows these equations to be further simplified:

\begin{align}
\overline{\sigma}_x &= -\frac{2\pi G \rho_2^2 a^2}{5} \bigg[A_x - \frac{2}{3}\frac{\rho_1}{\rho_2}\bigg(3+\frac{M_2}{M_1}\bigg)\bigg(\frac{d}{R_1}\bigg)^{-3}\bigg], \\
\overline{\sigma}_y &= -\frac{2\pi G\rho_2^2b^2}{5}   \bigg[A_y - \frac{2}{3}\frac{\rho_1}{\rho_2}\bigg(\frac{M_2}{M_1}\bigg)\bigg(\frac{d}{R_1}\bigg)^{-3} \bigg],\\
\overline{\sigma}_z &= -\frac{2\pi G \rho_2^2 c^2}{5} \bigg[A_z + \frac{2}{3}\frac{\rho_1}{\rho_2}\bigg(\frac{d}{R_1}\bigg)^{-3}\bigg].\\
\end{align}

\begin{figure}
\centering
\includegraphics[width=\linewidth]{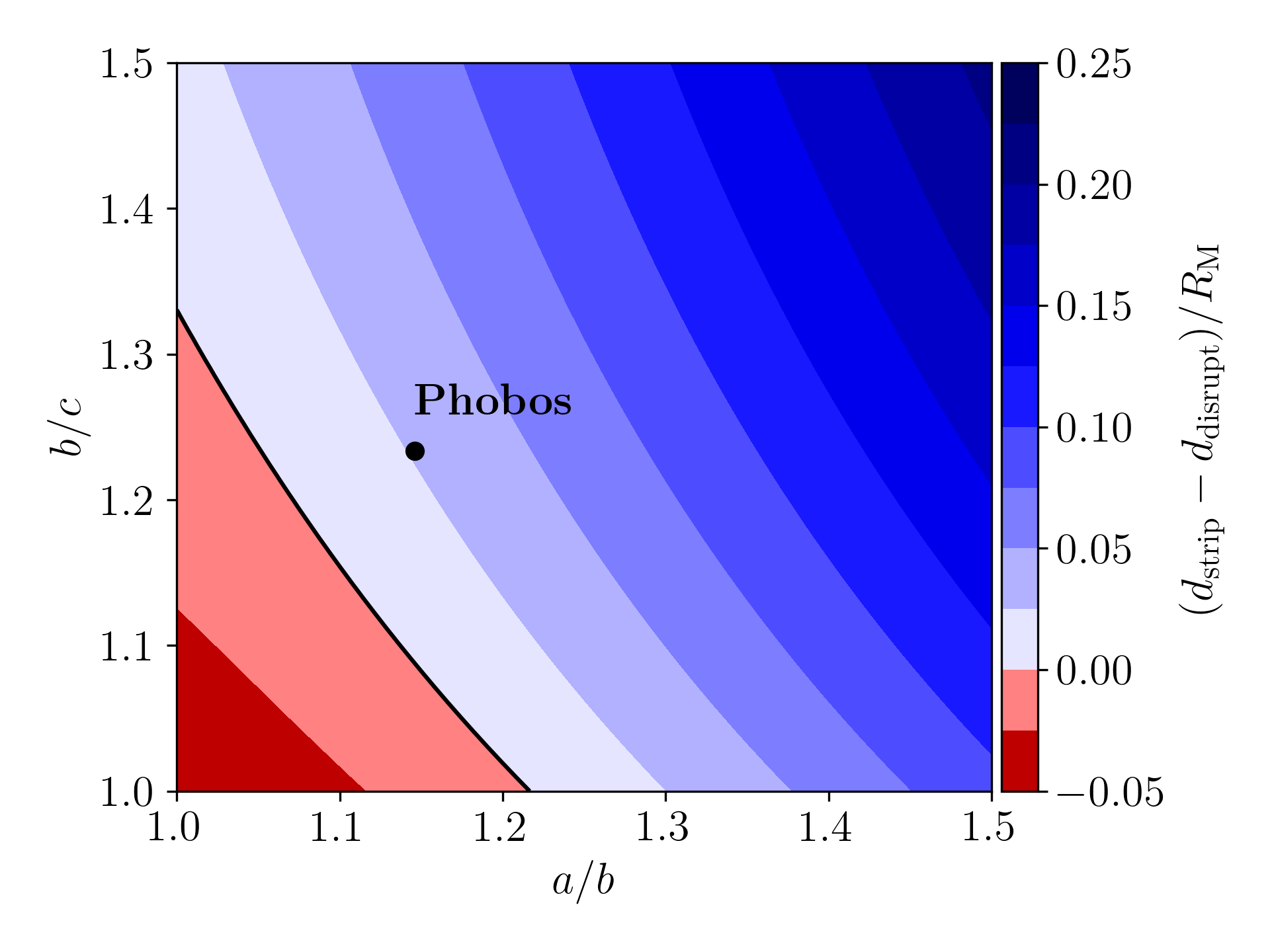}
\caption{\label{fig:disruption_map} Difference between the predicted distance for tidal stripping ($d_\text{strip}$) and the distance for disruption ($d_\text{disrupt}$), measured in Mars radii, as a function of a satellites axis ratios. When this difference is positive, tidal stripping is predicted to occur before internal failure (and tidal disruption). The satellite is assumed to have a triaxial shape, uniform density, a friction angle of $35^\circ$, and zero cohesion.}
\end{figure}

Using these equations, the critical distance, $d$, can be computed for a satellite of a given density, shape, friction, and cohesion at which the Drucker-Prager failure criterion is met and the satellite will undergo failure. Depending on the body's shape, this calculation can differ significantly from the simpler calculation of the rigid-body Roche limit in Sect. \ref{subsec:RRL}. {We note that this volume-averaging approach gives the orbital distance at which, on average, the failure criterion is violated \citep{Holsapple2007}. Therefore, this approach gives a lower limit on the tidal disruption distance, as failure can occur locally before the global, volume-averaged stress leads to failure \citep{Holsapple2008b,Holsapple2008}, meaning that some care should be taken when interpreting the derived Roche limits using this method. In the case of a cohesionless body, however, the volume-averaged approach gives an exact solution, because all locations within the body fail simultaneously \citep{Holsapple2006,Holsapple2008}. 

Given this caveat, we turn our attention to the special case when the cohesion is zero, and compare the disruption limit based on the Drucker-Prager yield criterion with the simpler approach that only considers surface accelerations.} A careful examination of these two calculations leads to an interesting conclusion: material can be stripped from an elongated object before it reaches the orbital distance for internal failure. In other words, the surface accelerations at the sub-primary point of a satellite can go to zero without the body stresses ever exceeding the failure criterion. In Fig.\ \ref{fig:disruption_map} we plot the distance between $d_\text{strip}$, the critical distance that allows for mass shedding (Eq. \ref{eq:d_shed}), and $d_\text{disrupt}$, the critical distance for disruption (Eq.\ref{eq:drucker}), for a Phobos-sized, cohesionless ellipsoid with a friction angle of $35^\circ$. This plot demonstrates that, in principle, an elongated rubble-pile object can be tidally stripped, while it migrates inward without undergoing a more violent disruption event. This effect is seen in the numerical simulations presented in the next section and we speculate that this could happen to Phobos, as long as it has a relatively low cohesion. This process is also general and could be applicable to any irregularly shaped satellite. This mechanism is qualitatively similar to the case of a differentiated satellite being stripped of its mantle owing to a density gradient \citep[e.g.,][]{Canup2010,Leinhardt2012b}.

\section{Numerical estimate for the Roche limit}\label{sec:numerical}
\subsection{Methods}\label{subsec:numerical_methods}
\begin{figure*}
\centering
\begin{subfigure}{0.33\textwidth}
  \centering
  \includegraphics[width=\linewidth]{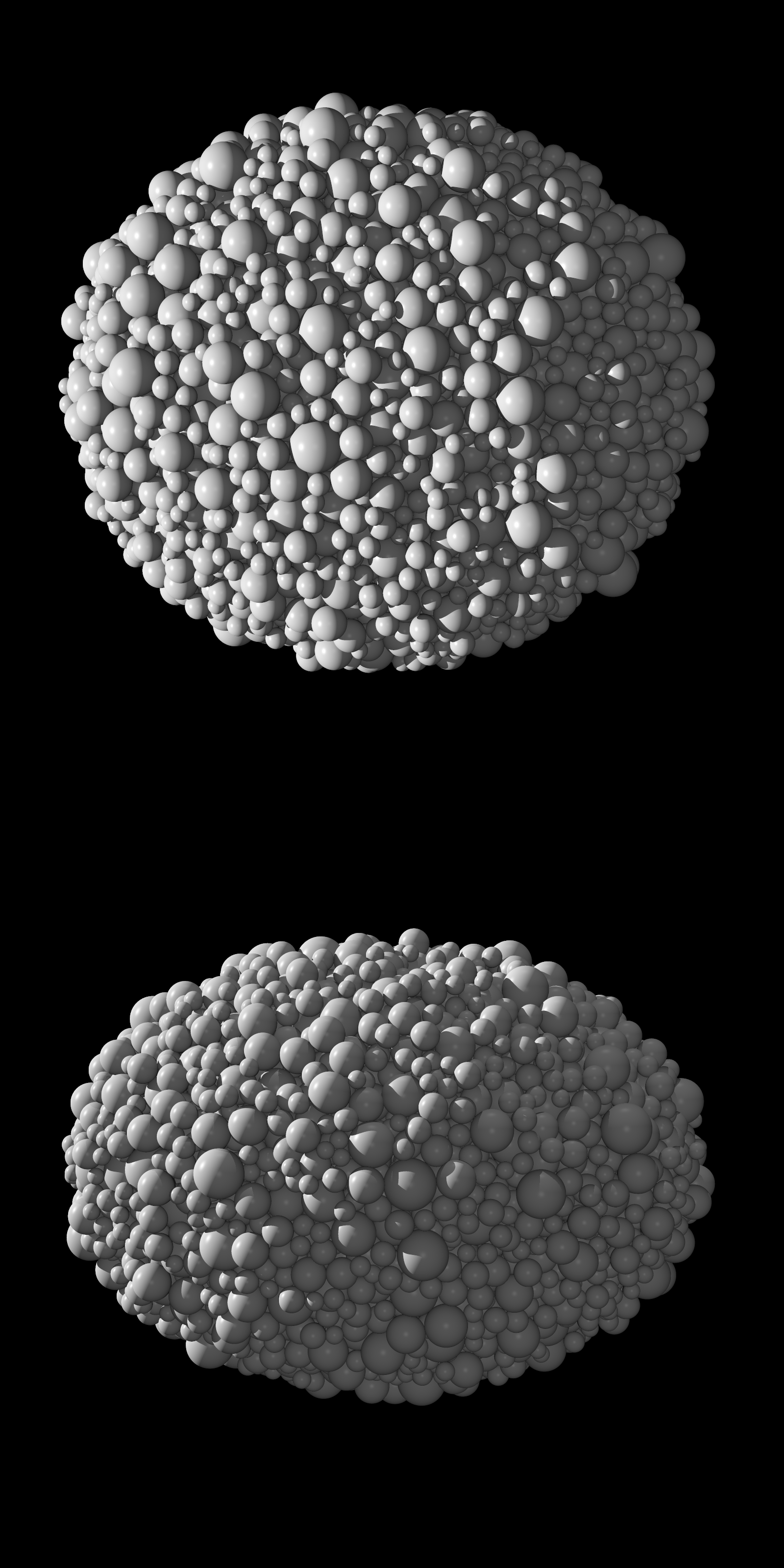}
  \caption{Ellipsoid, low resolution.}
  \label{subfig:rendering_ellipsoid}
\end{subfigure}\hfill
\begin{subfigure}{0.33\textwidth}
  \centering
  \includegraphics[width=\linewidth]{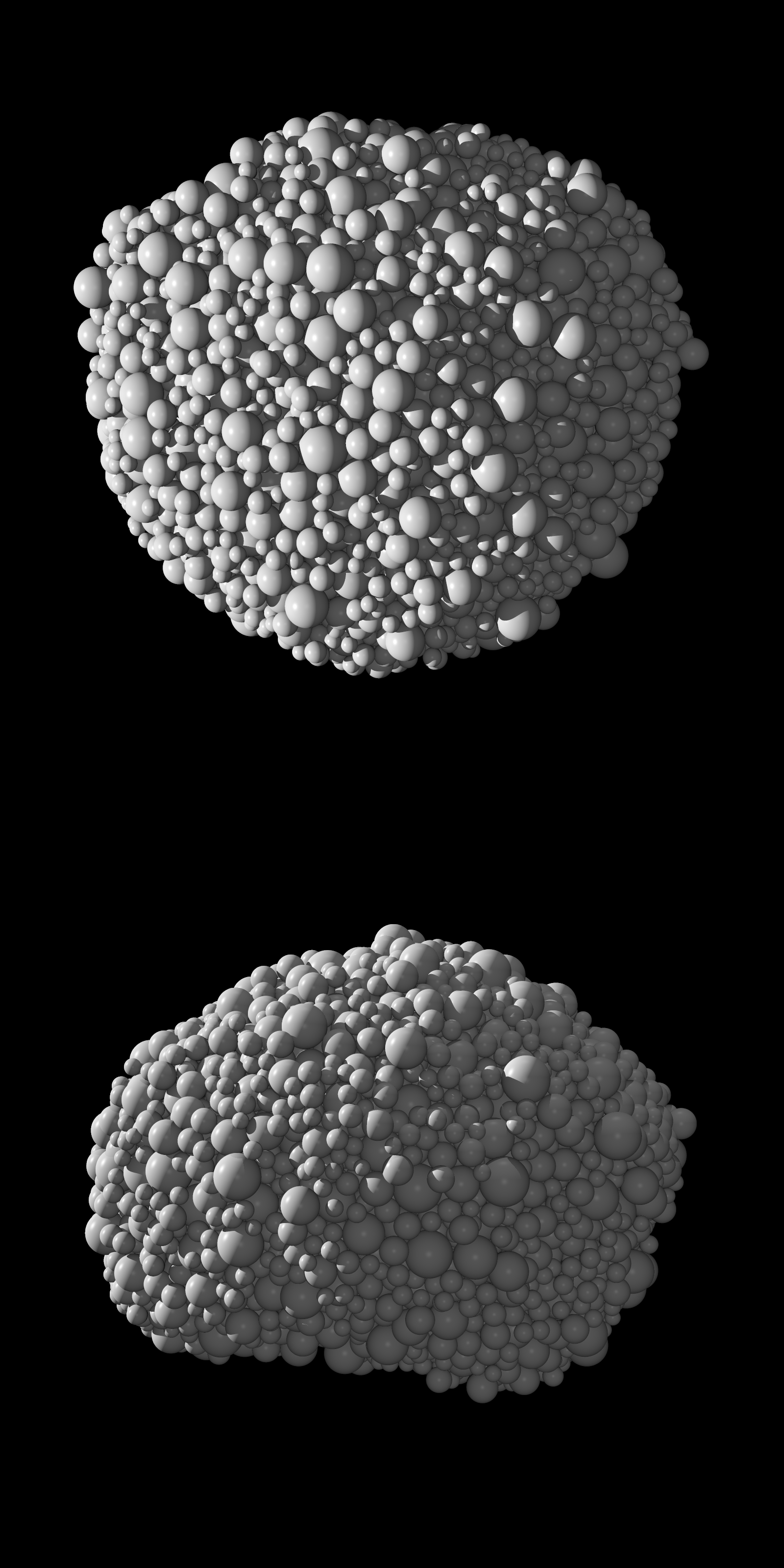}
  \caption{Phobos' shape, low resolution}.
  \label{subfig:rendering_crude}
\end{subfigure}\hfill
\begin{subfigure}{0.33\textwidth}
  \centering
  \includegraphics[width=\linewidth]{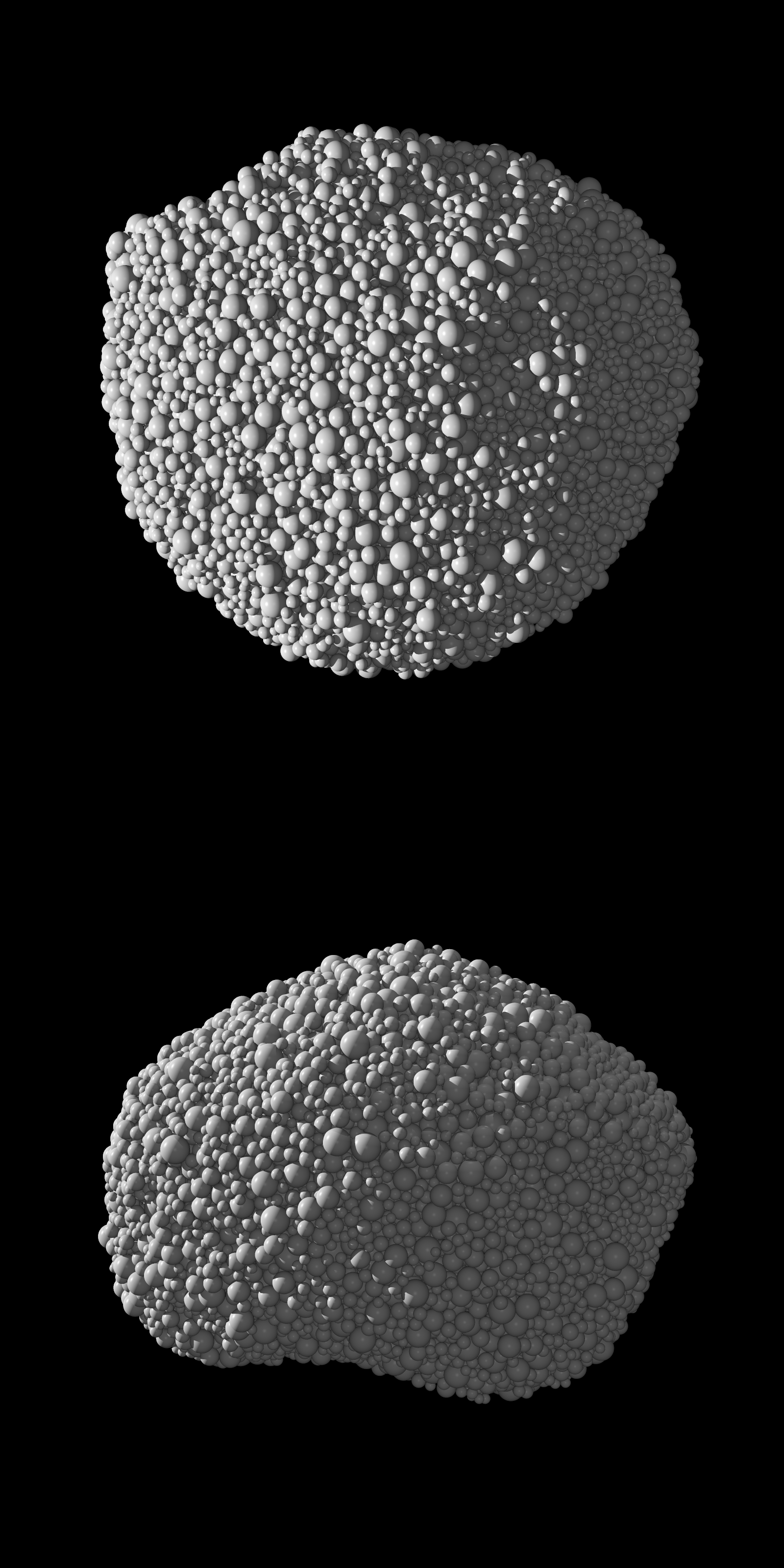}
  \caption{Phobos' shape, high resolution.}
  \label{subfig:rendering_high}
\end{subfigure}
\caption{Renderings of the rubble-pile models of Phobos used for this study. The top row shows a top-down rendering, looking down from Phobos' spin pole with Mars oriented to the left. The bottom row shows an edge-on view, where the camera is trailing Phobos orbit and Mars is to the left.}
\label{fig:renderings}
\end{figure*}

We used the gravitational $N$-body code \textsc{pkdgrav} along with its soft-sphere discrete element method (SSDEM) module to handle particle contacts and collisions \citep{Richardson2000,Schwartz2012,Zhang2017}. With SSDEM, the spherical particles are allowed to interpenetrate each other and feel a restoring force based on a damped Hookes law spring force. In addition, parameters for the particle's static, twisting, and rolling friction and restitution coefficients can be set so that the material will behave like a granular material. In this work, all SSDEM parameters were {chosen so that, in the absence of cohesion, the system behaves as a granular material with an angle of repose of ${\sim}35^\circ$, consistent with} values estimated for dry sand and for other small bodies \citep{Bareither2008,Fujiwara2006, Watanabe2019, Barnouin2019,Barnouin2024a,Robin2024a}. All simulations here consider a particle size-frequency distribution (SFD) with a differential power-law index of -3 that is comparable to the estimated boulder SFDs of rubble-pile asteroids visited by spacecraft \citep[e.g.,][]{Dellagiustina2019,Michikami2021,Pajola2024a}. We consider three realizations for a rubble-pile Phobos: a low-resolution, ellipsoid-shaped Phobos along with low- and high-resolution models using the real shape of Phobos. Due to computational constraints, the size ratio between the largest and smallest particle is set to a factor of 3. For the low-resolution simulations, Phobos consists of ${\sim}5500$ particles with a minimum particle radius of ${\sim}333$ m, while the high-resolution simulation has ${\sim}25000$ particles and a minimum particle radius of ${200}$ m. Renderings of the three models for Phobos are shown in Fig.\ \ref{fig:renderings}.

As a proxy for Phobos' tidal evolution, we artificially shrank its orbit in small increments{, such that each change in the orbital distance was very small compared to the orbit itself}. Starting at Phobos' present-day semimajor axis (${\sim}2.76R_\text{M}$), the satellite was integrated for a single orbital period before the orbit was shrunk by one Phobos diameter ($24.6\text{ km}{\sim}0.007R_\text{M}$), and the simulation restarted. At each restart, the satellite spin {rate was set to match the new mean motion} and its orbit was always circular. This process continued until the satellite reached $2.25R_\text{M}$, at which point its orbit was shrunk in increments of one Phobos radius rather than one diameter. After each orbit, Phobos was checked to see if it had lost any particles or its moments of inertia had changed by more than 1\%. If either of these conditions were met, the orbit shrinking process was paused and the simulation continued for ten additional orbital periods. The orbit shrinking process was only resumed after Phobos went ten consecutive orbits without the number of particles changing or the moments of inertia changing by 1\%. Any particles that escaped Phobos entered a circumplanetary orbit and had the possibility to re-impact or re-accrete on Phobos at a later time, although this was very rare given the short timescale of the simulations.

Although not perfectly quasi-static, this gradual orbit shrinking process allowed us to numerically determine the Roche limit for Phobos as a function of its material properties and internal structure as well as understand the mechanical behavior of a rubble pile as it approaches this limit. {These simulations migrate Phobos from its current orbit to its disruption limit on timescales of several tens of days, far shorter than its expected tidal evolution timescale of tens of millions of years \citep{Black2015}, which would be impossible to simulate numerically with thousands of particles. However, the mass stripping events seen in the simulations occur on timescales of hours, which are well captured in these simulations.}

We considered a model for the cohesive force between particles of the form
\begin{equation}
    F_c=cA_\text{eff},   
\end{equation}
where $A_\text{eff}$ is the effective contact area between two particles and $c$ is an interparticle cohesive constant \citep{Zhang2018,Zhang2021}. The contact area is approximated by $A_\text{eff}=(2\beta r_\text{eff})^2$, where $\beta$ is a dimensionless shape parameter and $r_\text{eff}=r_i r_j/(r_i+r_j)$ is the effective contact radius of the two particles, $i$ and $j$, in contact having radii $r_i$ and $r_j$.  

We considered three different values for the interparticle cohesive constant c: $0,10^4,10^6$ Pa. The bulk cohesive strength can be related to the interparticle cohesive strength by
\begin{equation} \label{eq:C}
    C\sim\frac{\beta^2N_\text{c}\eta\tan\phi}{4\pi}c,
\end{equation}
where $N_\text{c}$ is the average number of contacts for each particle and $\eta$ is the packing efficiency of the rubble pile \citep{Rumpf1958, Zhang2021}. This expression was derived assuming that all particle contacts in an assembly break simultaneously, meaning that $C$ should be interpreted as an upper limit of a given body's bulk cohesive strength. For the rubble pile models under consideration, the average packing fraction and contact number are ${\sim}0.6$ and ${\sim}5$, respectively, and all simulations use $\beta=0.5$, meaning that the bulk cohesive strength, $C$, is approximately $4\%$ of the interparticle cohesive constant. The three bulk cohesive strengths under consideration are then 0, 0.4, and 40 kPa. {The largest cohesion value considered is comparable to the interior overburden pressure at the center of Phobos, which is ${\sim}60$ kPa, if Phobos were a uniform density sphere.}

\subsection{Results}

\begin{figure*}
\centering
\begin{subfigure}{0.5\linewidth}
  \centering
  \includegraphics[width=\linewidth]{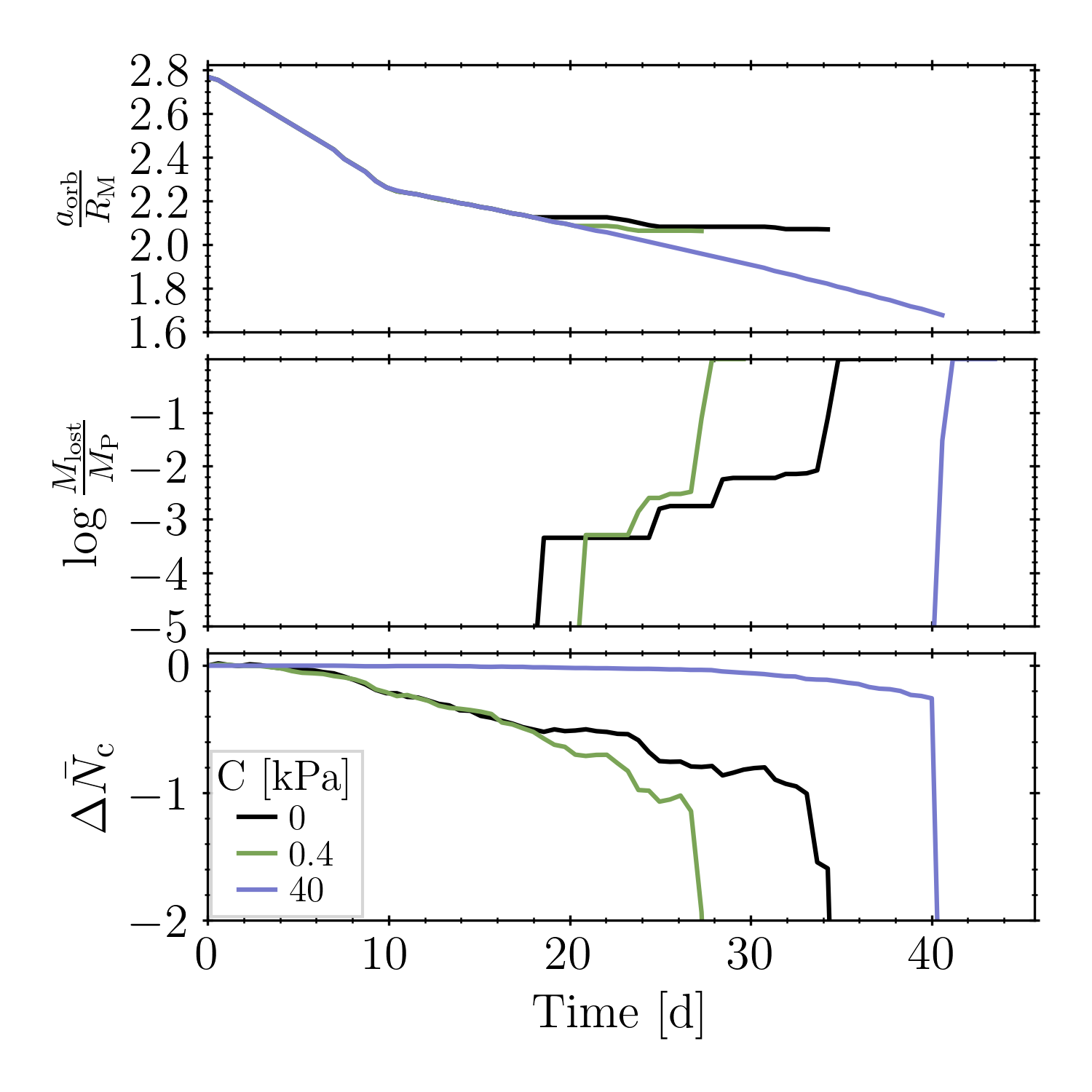}
  \caption{Ellipsoid, low resolution.}
  \label{subfig:timeseries_ellipsoid}
\end{subfigure}\hfill
\begin{subfigure}{0.5\linewidth}
  \centering
  \includegraphics[width=\linewidth]{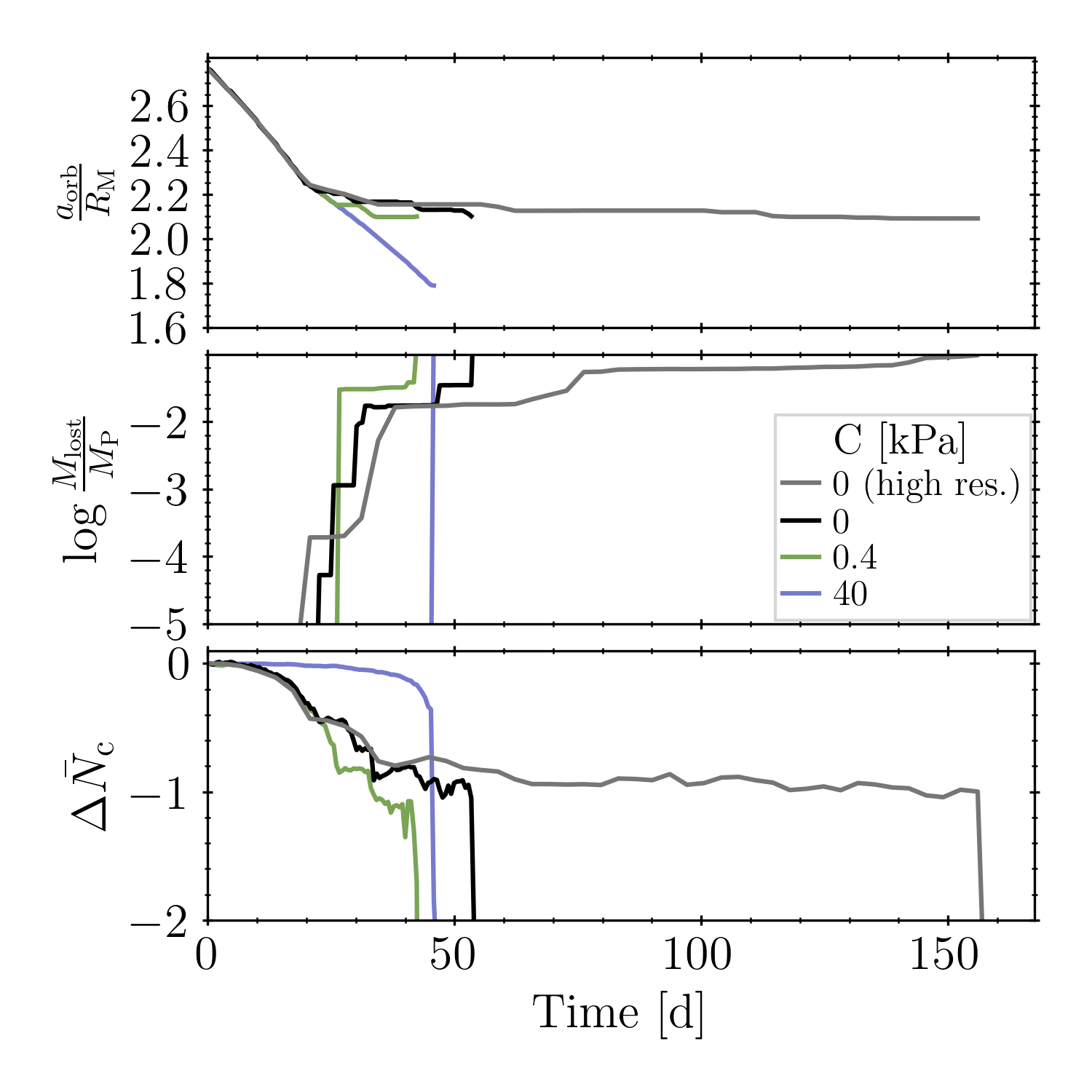}
  \caption{Phobos' shape, low and high resolutions.}
  \label{subfig:timeseries_crude}
\end{subfigure}%
\caption{Time-series plots showing the semimajor axis ($a_\text{orb})$, the amount of lost mass ($M_\text{lost}/M_\text{P}$), and the change in the average contact number ($\Delta\bar{N}_c$) for the three rubble pile models of Phobos for different bulk cohesive strengths. The ellipsoidal case is shown in panel (a), while the low- and high-resolution cases with the real Phobos shape are combined in panel (b).  \label{fig:timeseries}}
\end{figure*}

In Fig.\ \ref{fig:timeseries}, we show the time history of Phobos' semimajor axis ($a_\text{orb}$), the mass lost ($M_\text{lost}$), and the change in average contact number ($\Delta \bar{N}_c$) for the three rubble pile models. We see that the low cohesion cases undergo several discrete mass shedding events as they migrate inward before they are ultimately disrupted. This behavior is a result of the net gravitational acceleration going to zero at the sub- and anti-Mars point before the body undergoes internal failure as expected based on Fig.\ \ref{fig:disruption_map}. However, when the cohesion is sufficiently high, mass shedding is impossible and the body migrates much closer to Mars before suddenly disrupting as a result of internal failure. If our orbital evolution were more gradual and occurred over realistic timescales, we speculate that a weak Phobos could be destroyed purely by tidal stripping without ever having an abrupt tidal disruption. However, this is not computationally feasible to test.

In Fig.\ \ref{fig:snapshots_highres}, we show several snapshots of the high-resolution Phobos case to qualitatively demonstrate the mass shedding events. In these renderings, the colors of each particle indicate the magnitude of each particles net acceleration (the vector sum of self-gravity, tidal, and centrifugal forces) with respect to Phobos' center of mass. This demonstrates that near the sub- and anti-Mars points, the net acceleration vanishes, allowing material to be freely stripped from the surface. The first shedding event occurs after ${\sim}20$ d at a distance of ${\sim}2.25\RM$, where several particles are lost from an area near the rim of Stickney crater. Several larger mass shedding events occur at ${\sim}2.15\RM$ and ${\sim}2.13\RM$. Finally, at ${\sim}2.09\RM$, Phobos undergoes a large shedding event, which also provides a torque to the remaining mass, destabilizing its synchronous rotation state and leading to complete tidal disruption.

\begin{figure*}
\centering
\includegraphics[width=0.44\textwidth]{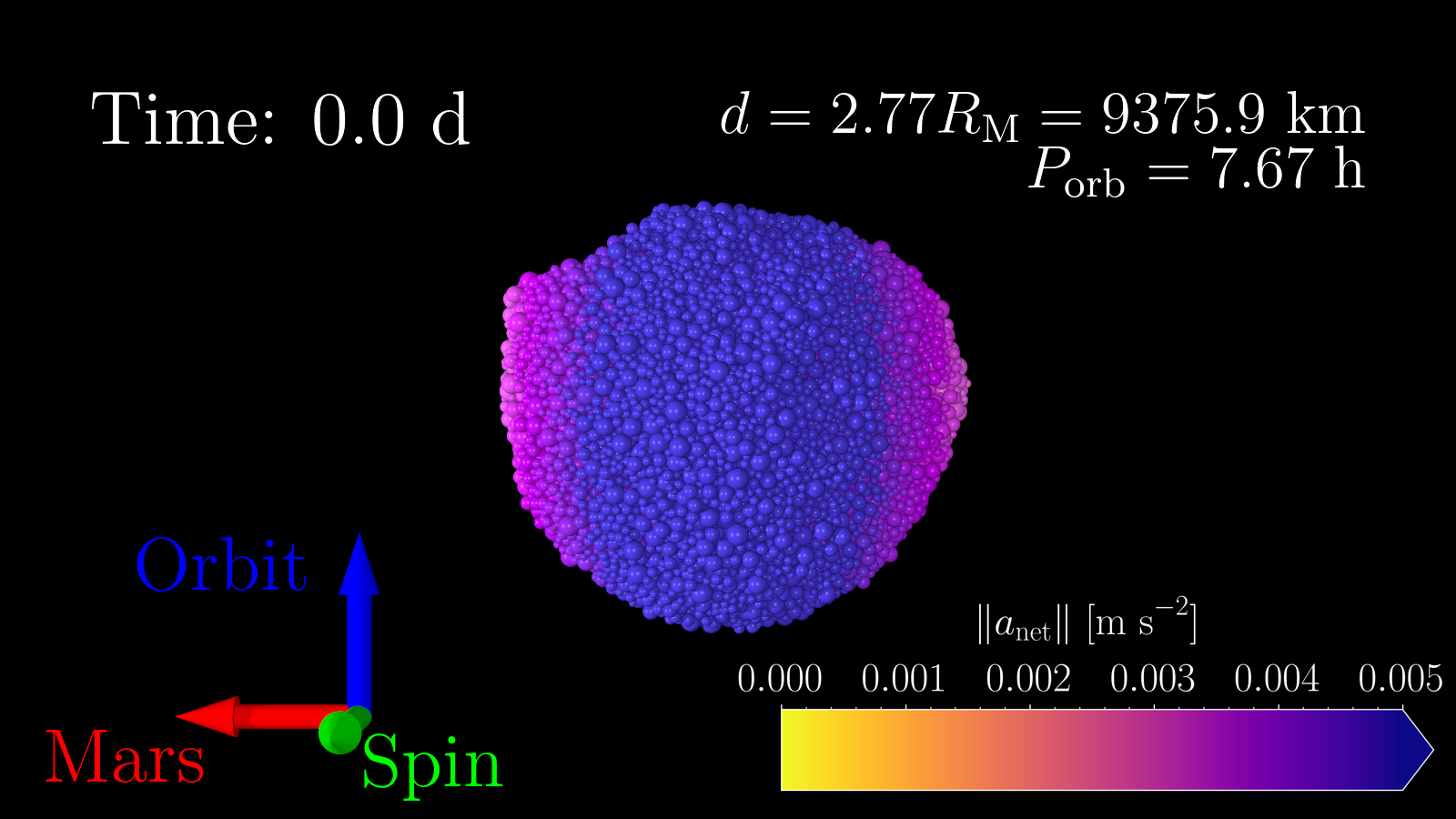}\hspace{0.5mm}
\includegraphics[width=0.44\textwidth]{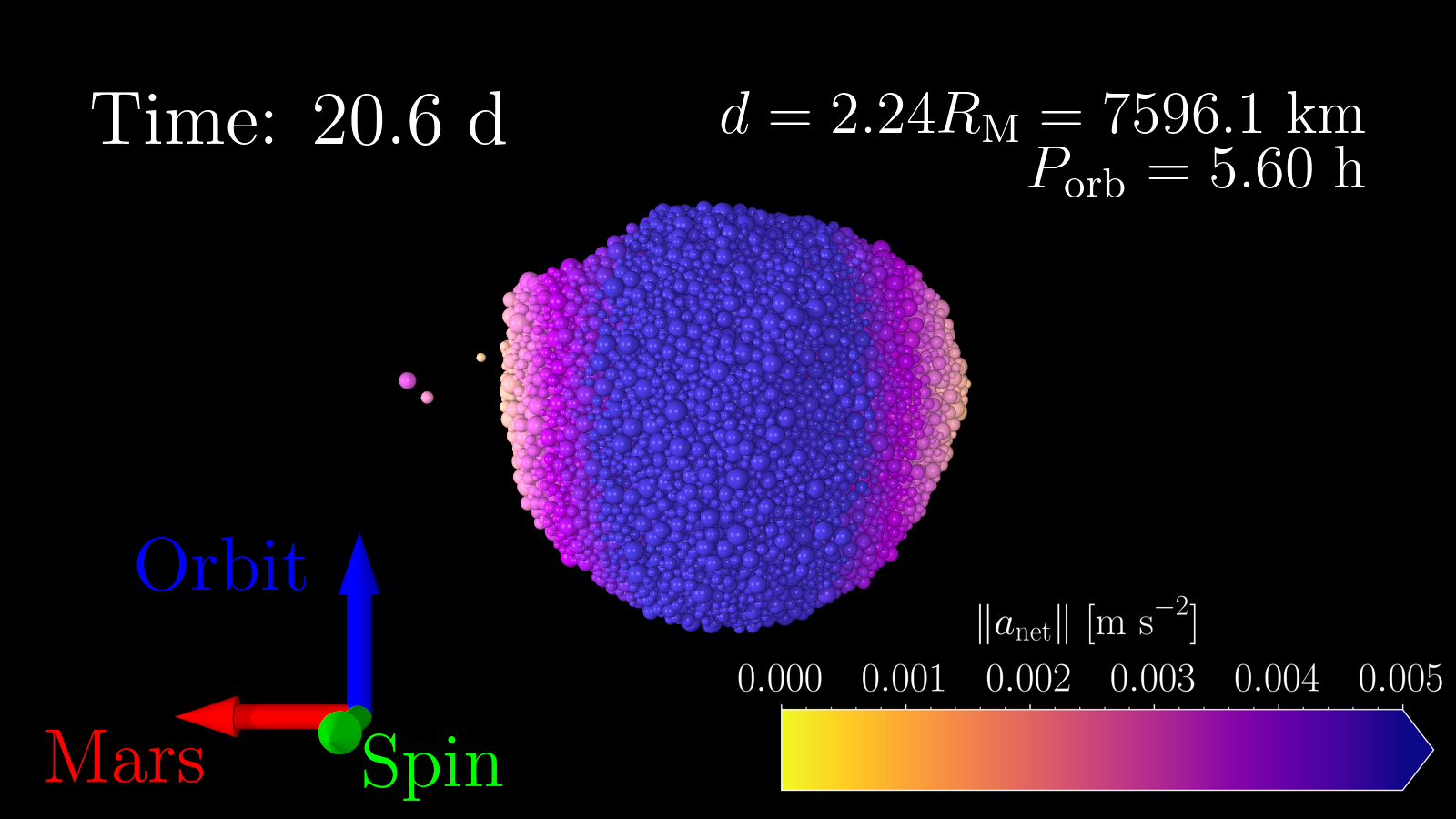}

\vspace{0.5mm}

\includegraphics[width=0.44\textwidth]{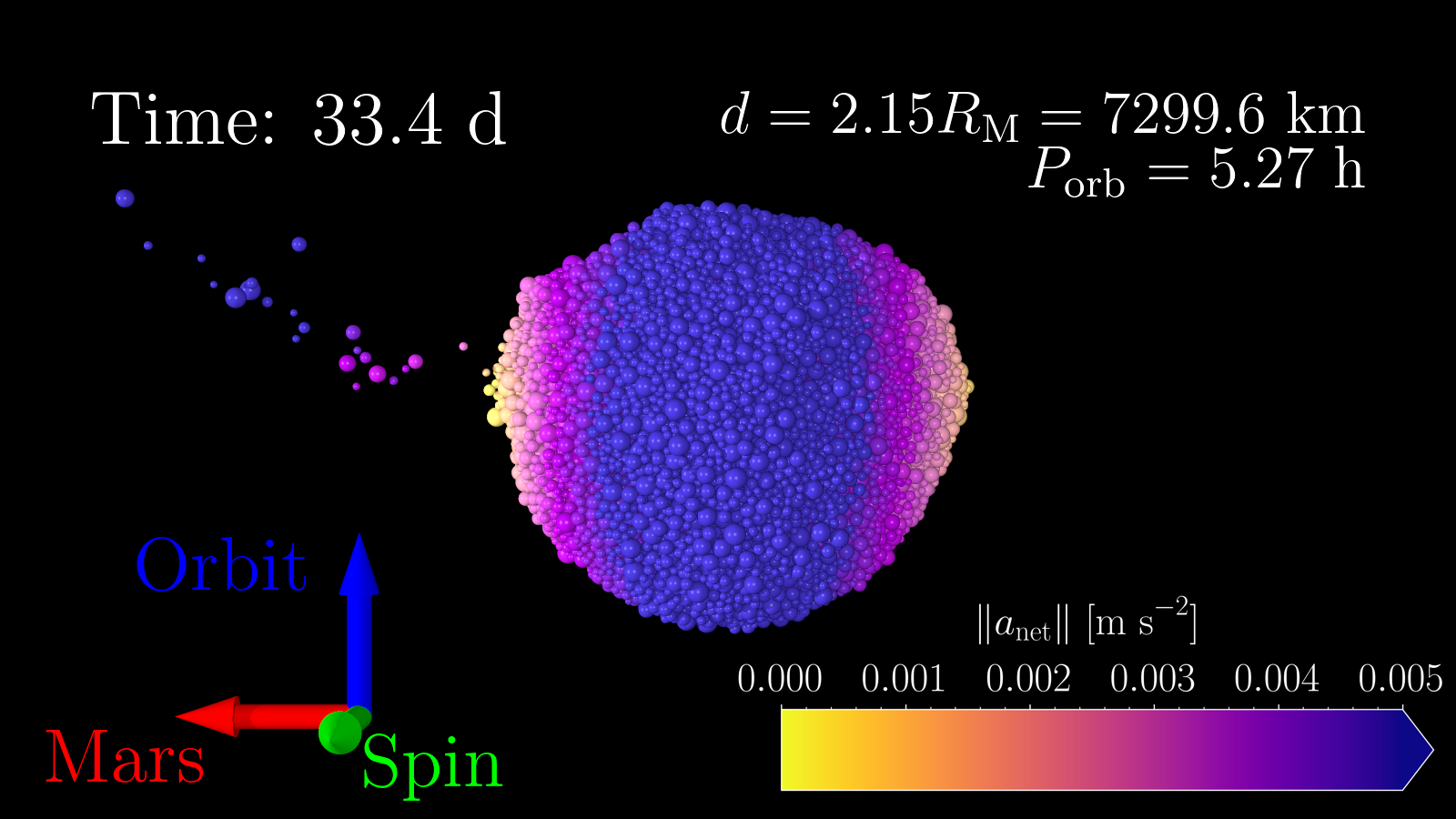}\hspace{0.5mm}
\includegraphics[width=0.44\textwidth]{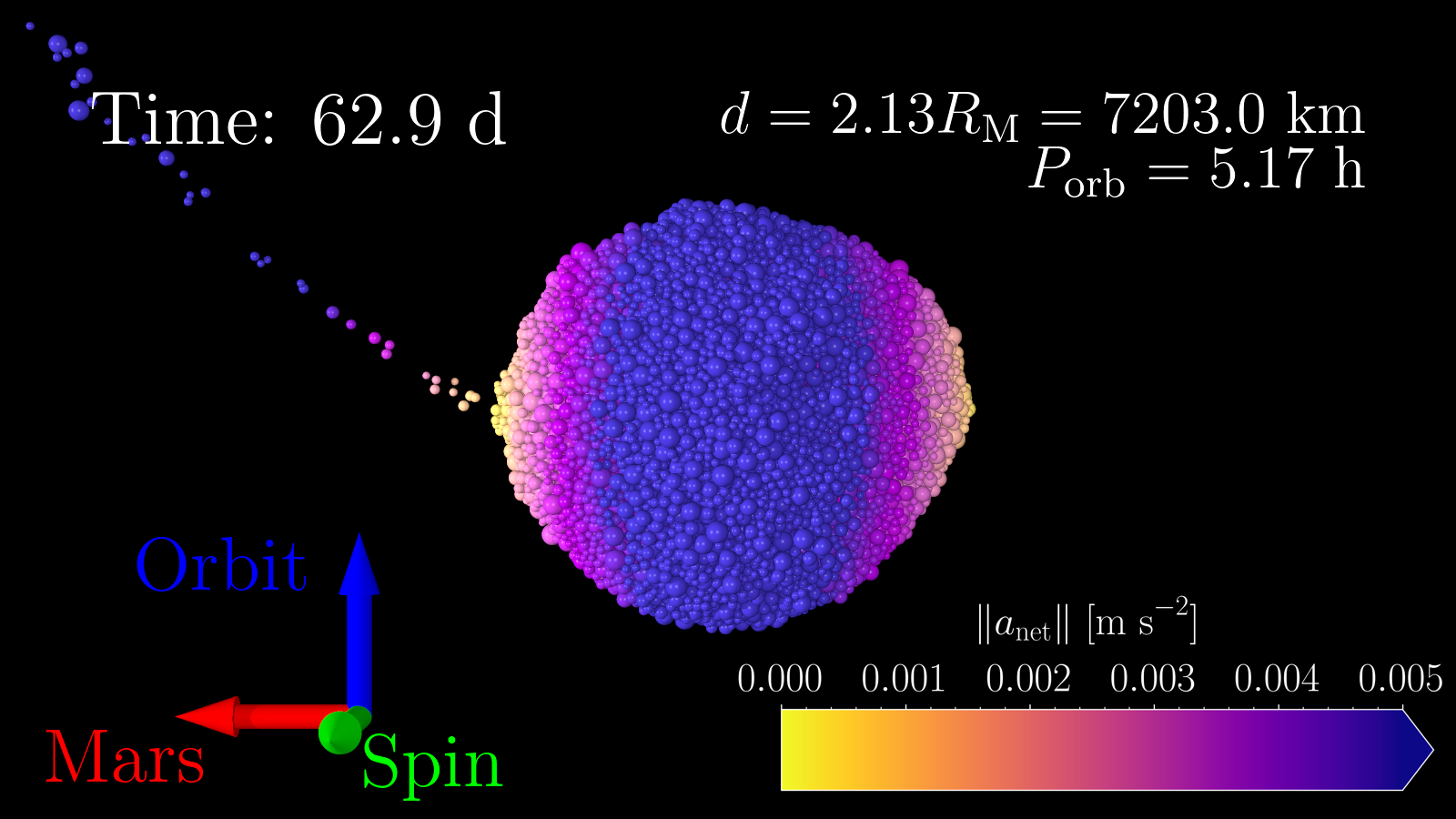}

\vspace{0.5mm}

\includegraphics[width=0.44\textwidth]{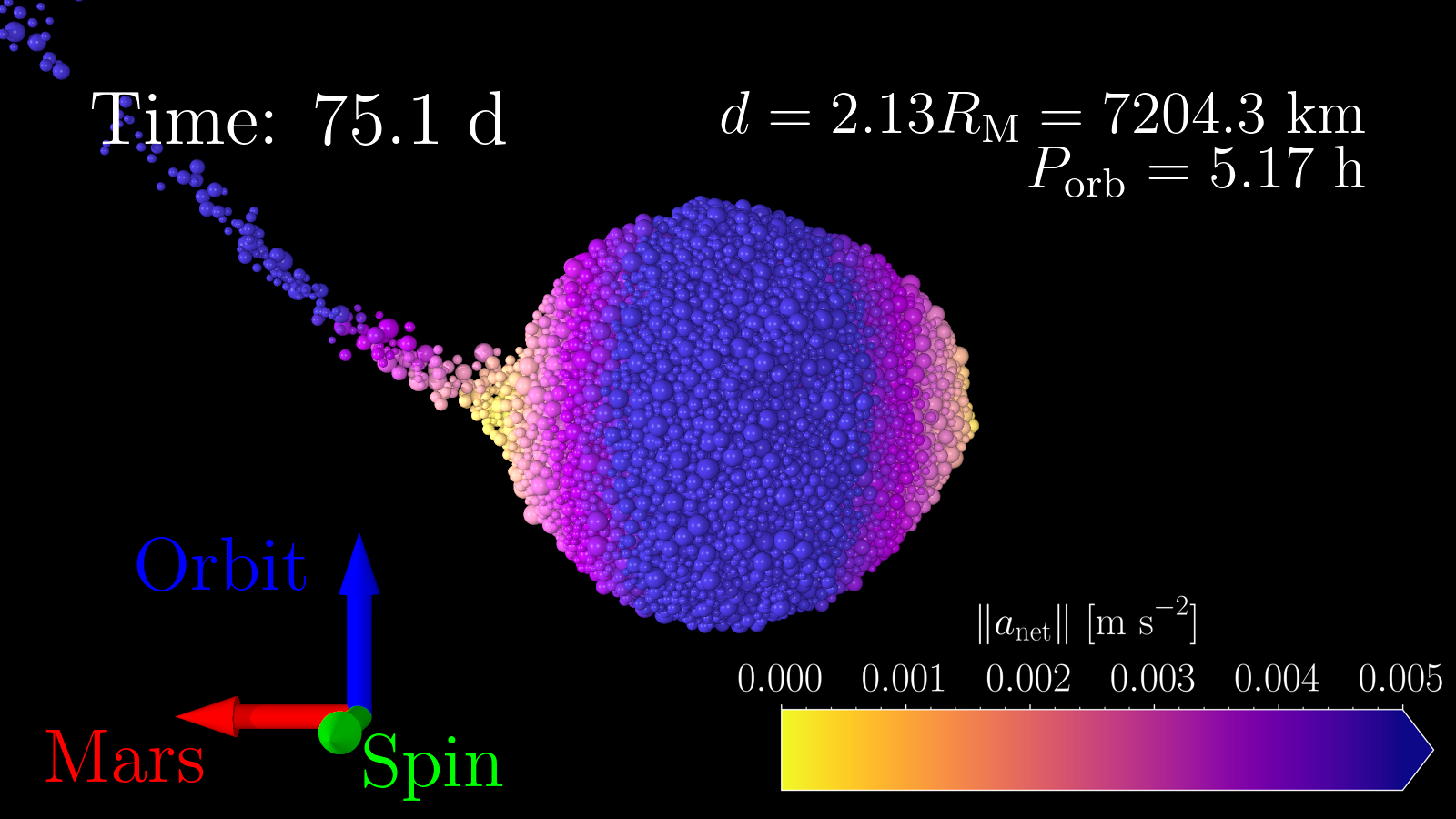}\hspace{0.5mm}
\includegraphics[width=0.44\textwidth]{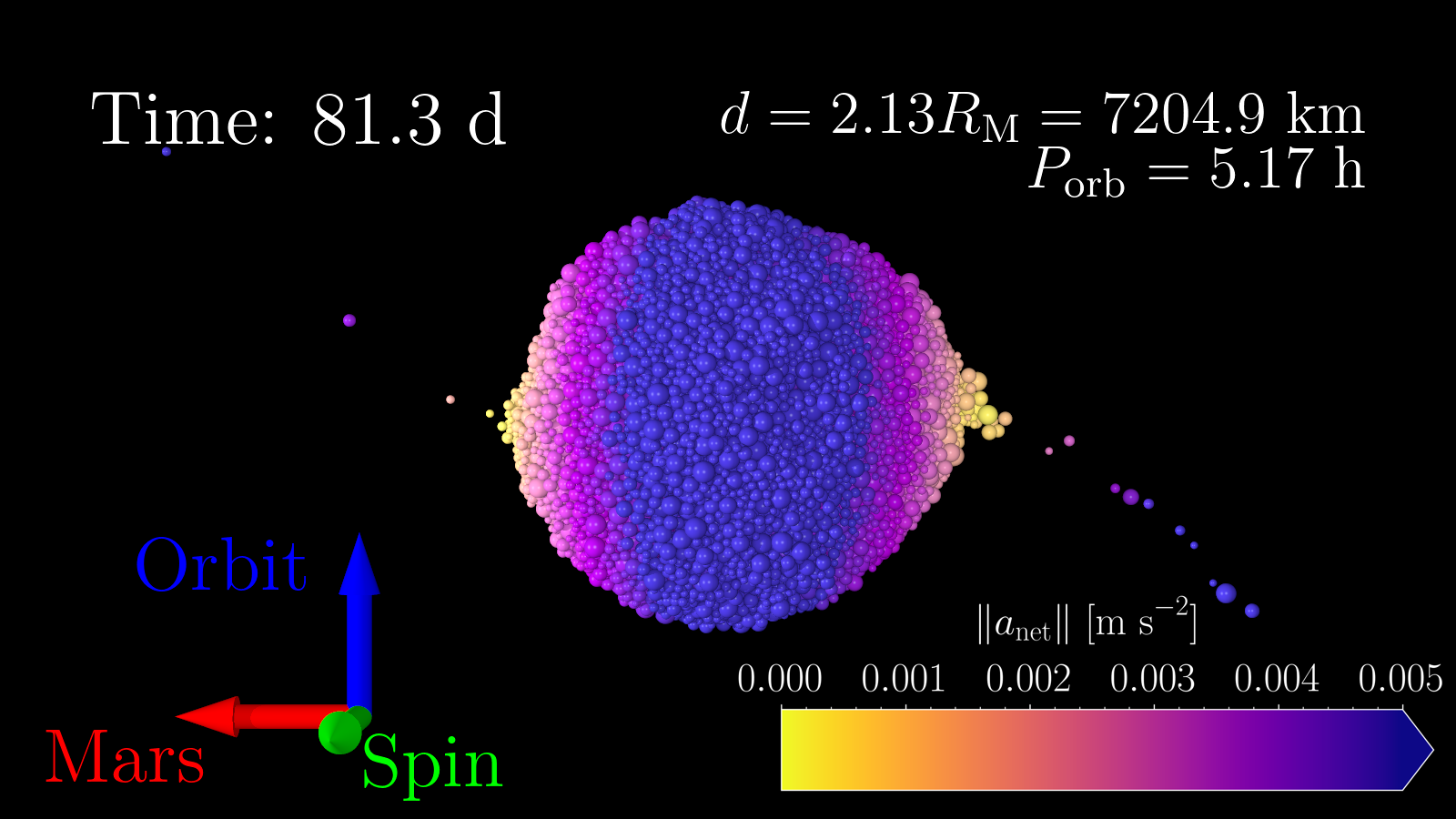}

\vspace{0.5mm}

\includegraphics[width=0.44\textwidth]{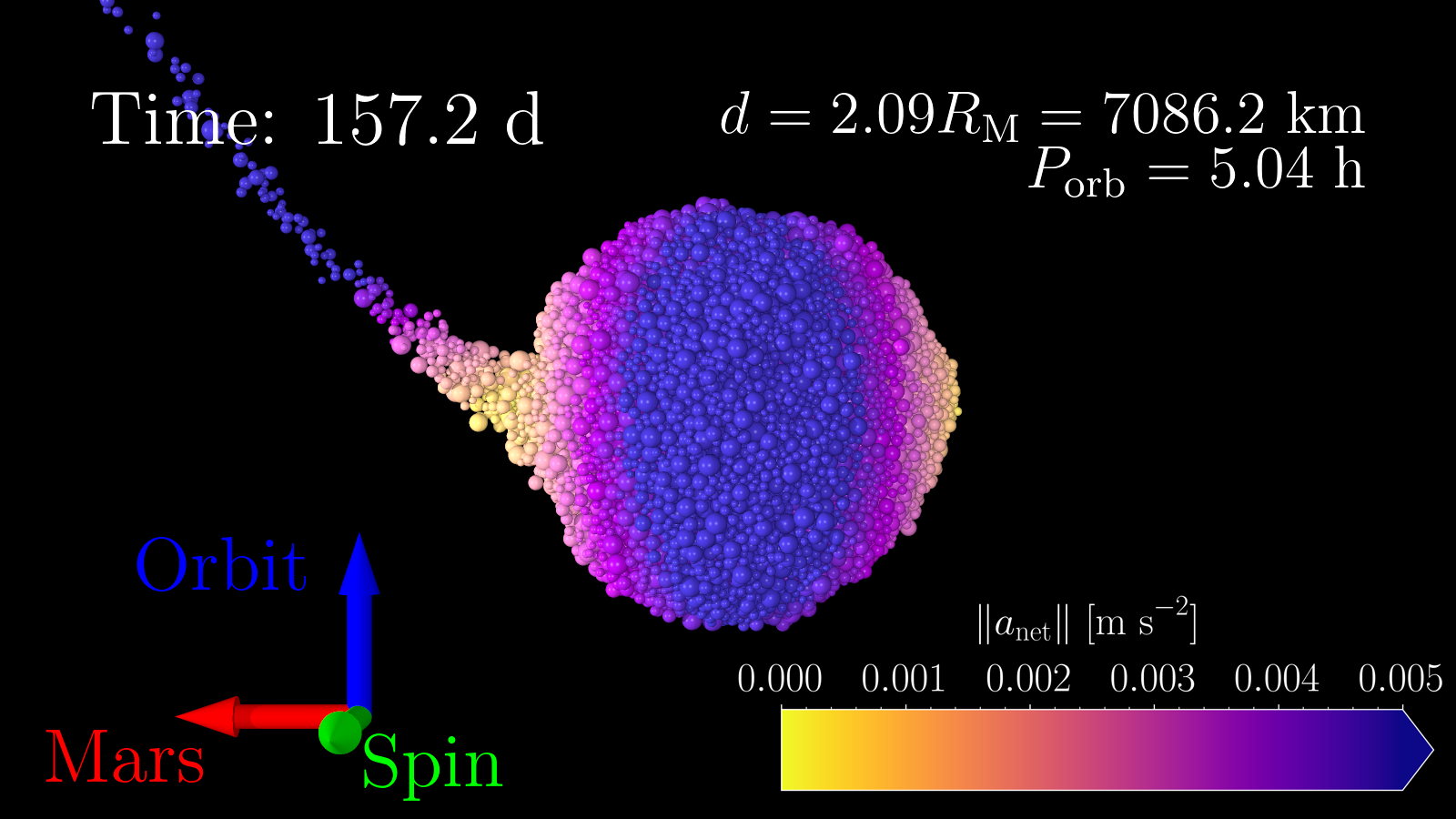}\hspace{0.5mm}
\includegraphics[width=0.44\textwidth]{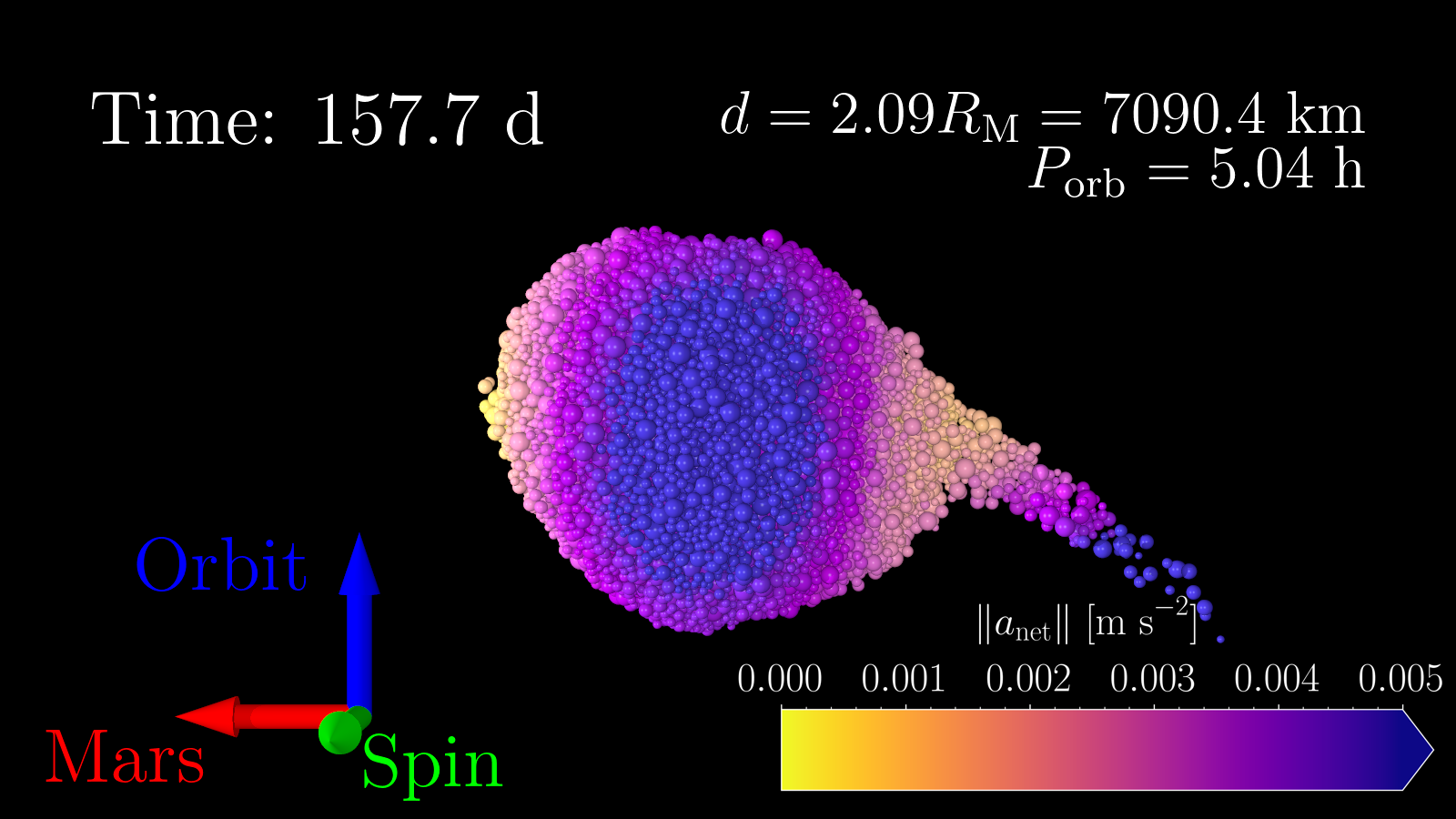}

\vspace{0.5mm}

\includegraphics[width=0.44\textwidth]{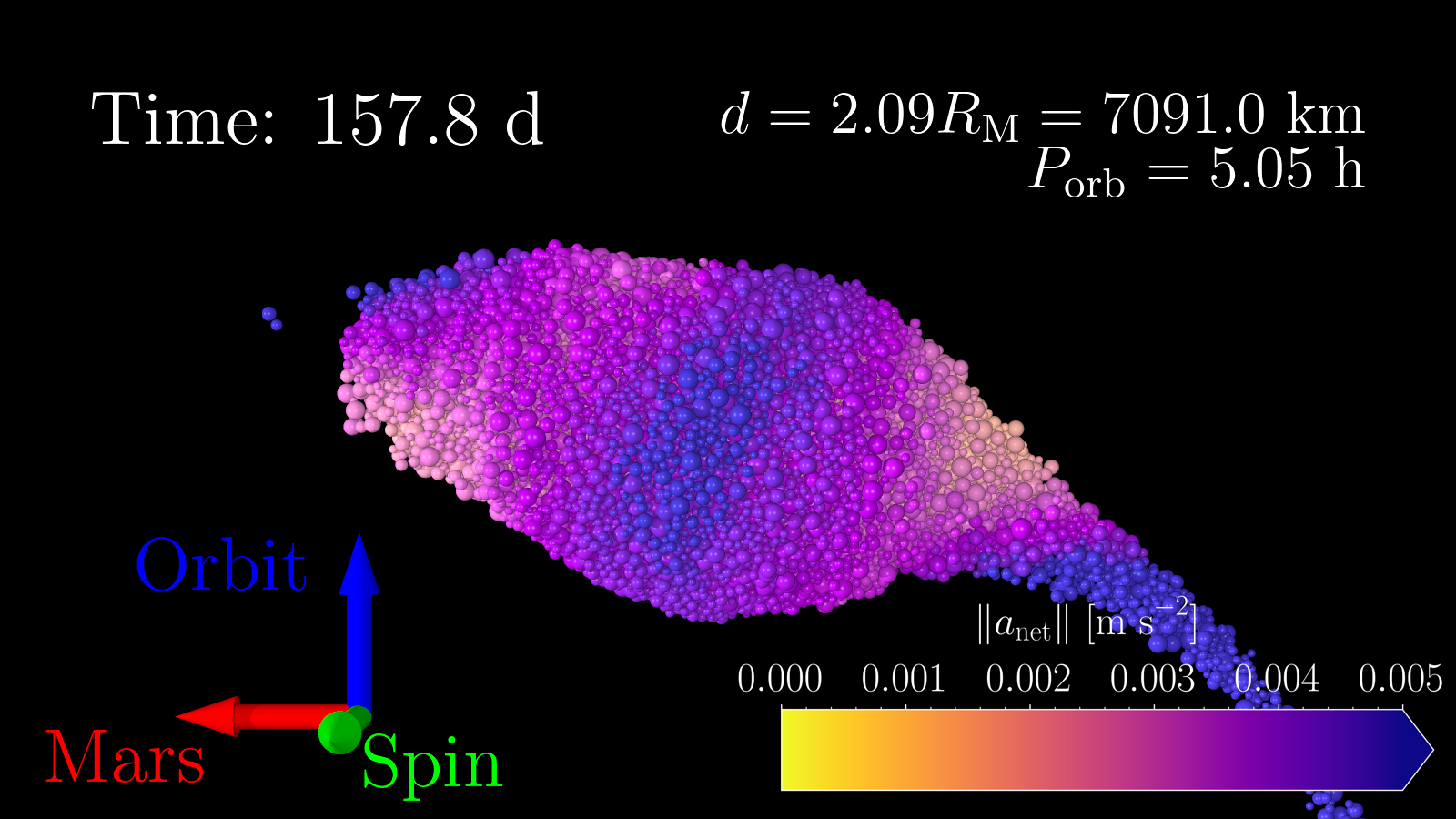}\hspace{0.5mm}
\includegraphics[width=0.44\textwidth]{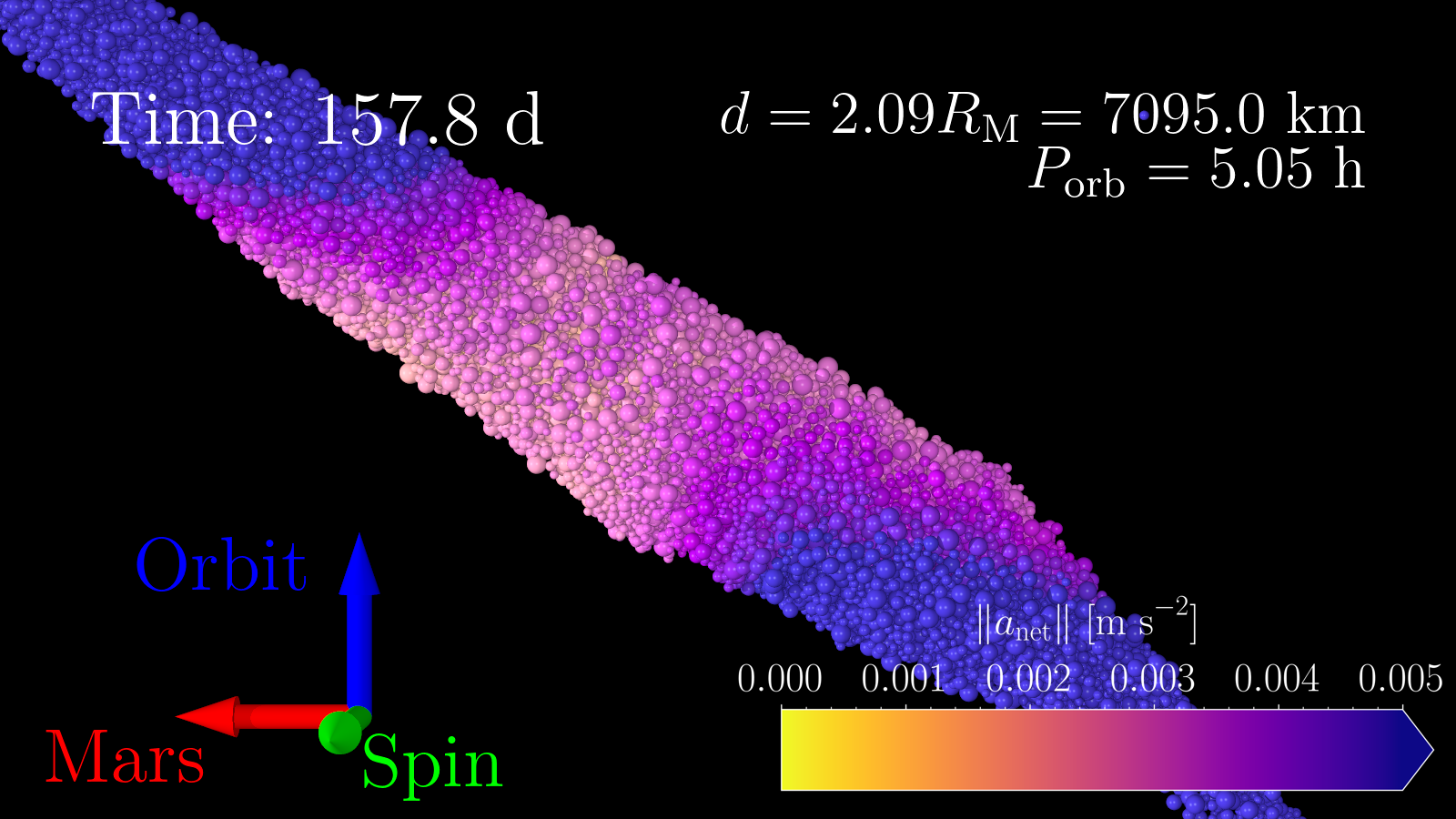}

\caption{\label{fig:snapshots_highres} Snapshots of the cohesionless, high-resolution Phobos case, in a coorbiting frame looking down from Phobos spin pole with Mars oriented to the left. Each frame indicates the simulation time, orbital distance, and orbital period (which equals Phobos spin period). The particle colors indicate their net acceleration magnitude relative to Phobos' center of mass. Several mass shedding events occur starting at ${\sim}2.25\RM$ until Phobos is finally disrupted at ${\sim}2.09\RM$. There are many mass shedding events in addition to the three shown in this figure and we refer the reader to the {animated version of this figure.}}
\label{fig:snapshots_highres}
\end{figure*}

\begin{figure}
    \centering
    \includegraphics[width=\linewidth]{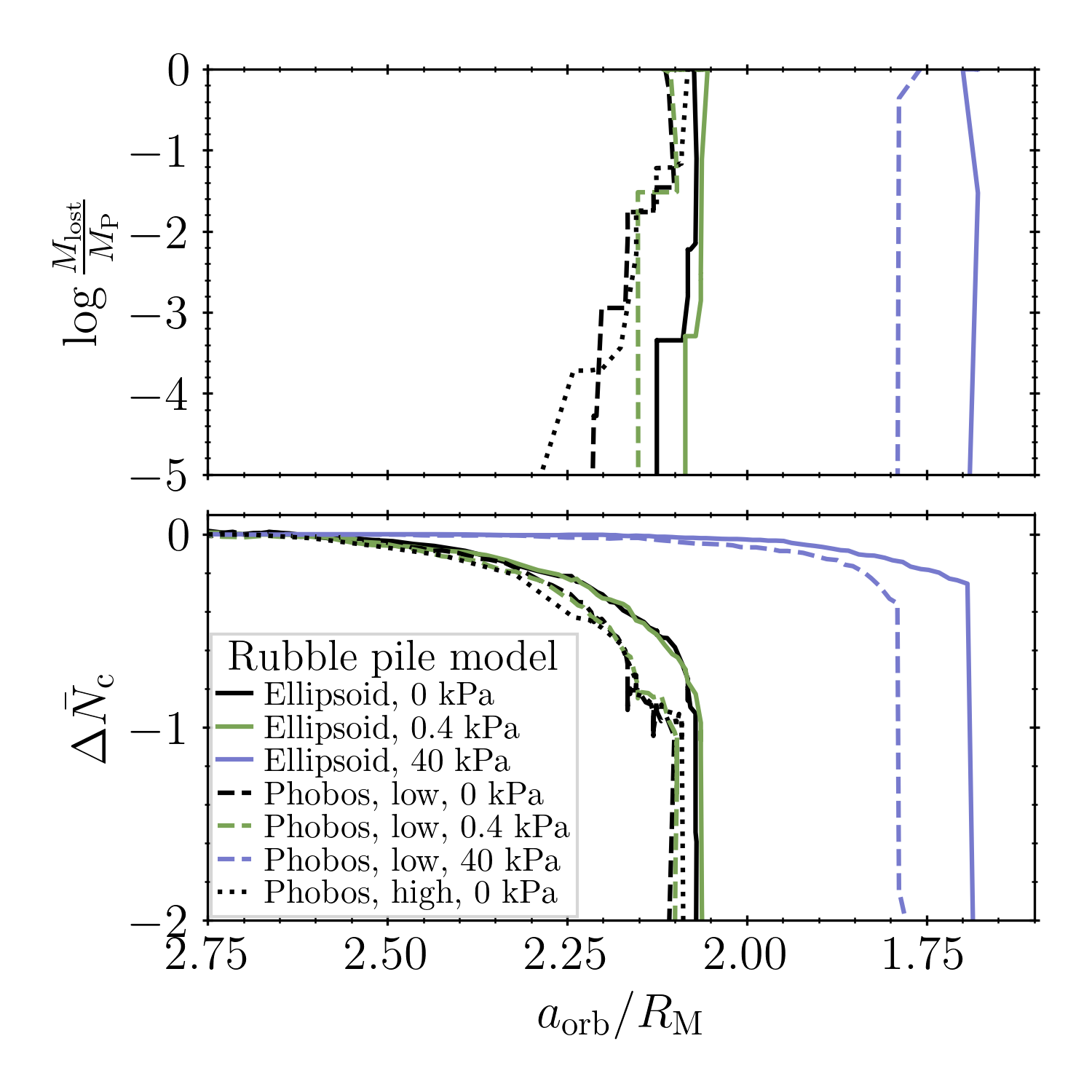}
    \caption{\label{fig:a_vs_M_C} Total mass lost and change in mean contact number as a function of the satellite's orbital distance for all simulations. The style of each line (solid, dashed, dotted) indicates the rubble pile model, while the color of each line indicates the cohesive strength.  }
\end{figure}

As a result of the artificial migration procedure being adaptive, the different rubble pile models disrupt at different times, which makes it difficult to compare them directly. It is more useful to plot the mass lost and contact number as a function of the orbital distance rather than time, which is shown in Fig. \ref{fig:a_vs_M_C}. One striking feature in these plots, especially for the cases with low cohesion, is that the average contact number drops significantly as Phobos migrates inward. For many cases, $\bar{N}_\text{c}$ drops by ${\sim}1$ (or ${\sim}20\%$) before the body ultimately undergoes failure. Recall that the bulk cohesive strength depends linearly on the number of particle contacts. Effectively, this translates to a decrease in the bulk strength of ${\sim}20\%$, which means that the body ends up disrupting further out than what would be predicted by its initial bulk strength. {Some caution is warranted in interpreting this result, however, as these simulations employ a narrow particle size–frequency distribution. Consequently, they do not capture the potential role of fine-grained material that could percolate into void spaces, preserve the contact network, and help maintain the body's bulk cohesive strength.}

Figure \ref{fig:a_vs_M_C} shows that the two Phobos models (especially the high-resolution case) initiate mass shedding at a slightly larger distance than the ellipsoid case when the cohesion is small, owing to the body's different gravitational potential and higher initial surface slopes. Phobos' tidal disruption is relatively insensitive to cohesion at the level of hundreds of pascals, given the negligible differences between the 0 and 0.4 kPa cases for both the ellipsoid and low-resolution Phobos models. The cohesive strength only makes a significant difference at the level of several or tens of kilopascals, which is consistent with the semi-analytic work by \cite{Holsapple2008}. Finally, the high- and low-resolution Phobos cases show similar behavior, which gives some confidence that the other low-resolution simulations are sufficiently capturing the bulk behavior. {Even the high-resolution case for Phobos only marginally resolves high-slope regions such as those around Stickney crater. In these regions of high slopes, landslides can be triggered, mobilizing material that could make its way into orbit. Therefore, we suggest that mass stripping could occur at even greater distances than demonstrated here.}

\begin{figure}
\centering
\includegraphics[width=\linewidth]{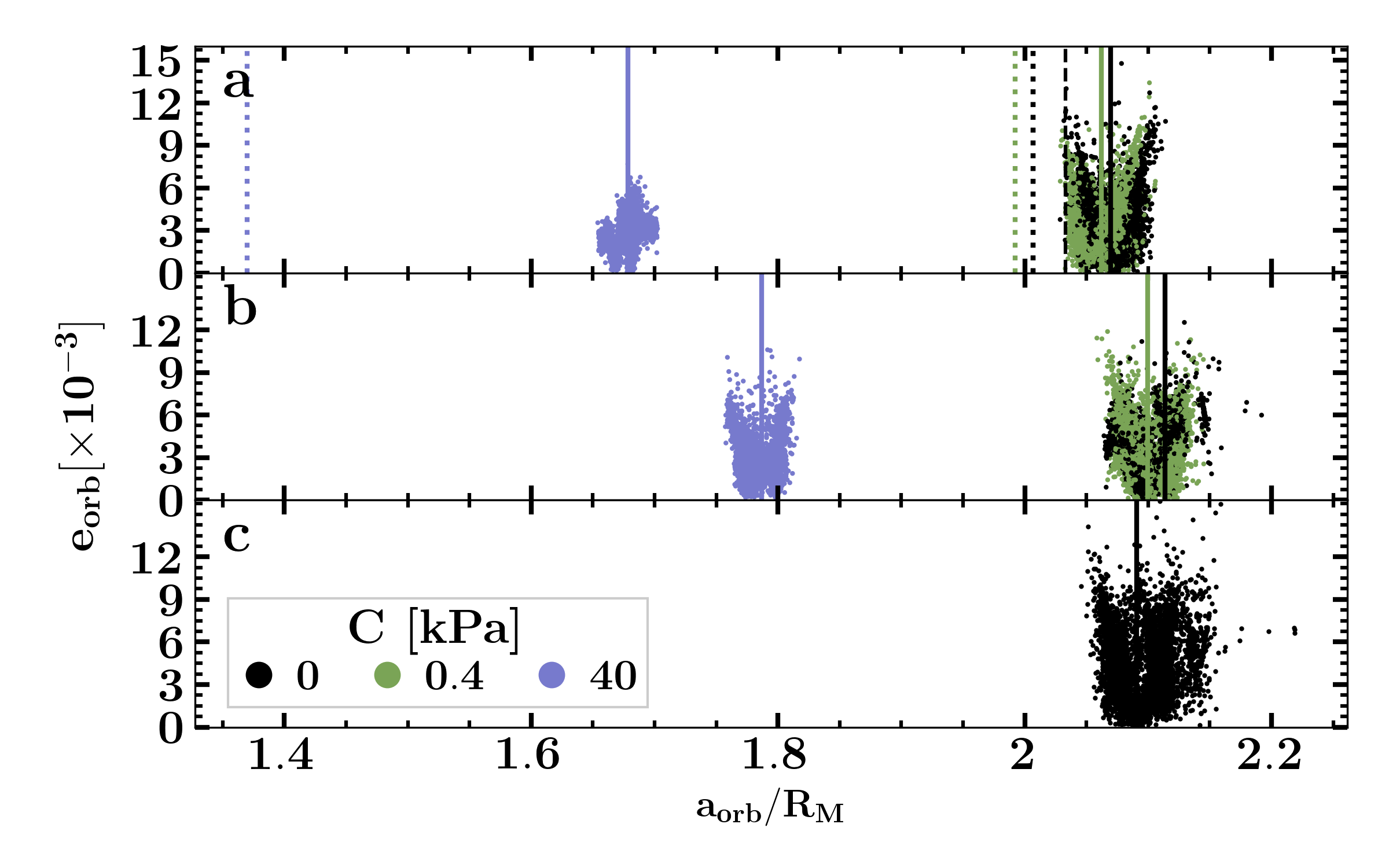}
\caption{\label{fig:orbElem} Orbital semimajor axes and eccentricities of all particles after Phobos is tidally disrupted. The solid vertical lines denote the distance at which Phobos was tidally disrupted, which we define as the point at which its mass is half of its initial value. (a) Low-resolution cases with an ellipsoidal Phobos. The dashed black line indicates the second-order rigid-body Roche limit derived in Sect. \ref{subsec:RRL} and the dotted lines indicated the predicted disruption limit based on the Drucker-Prager failure criterion derived in Sect. \ref{subsec:druckerprager}. (b) Low-resolution cases based on Phobos' true shape model. (c) Single high-resolution case using Phobos' shape model with zero cohesion.}
\end{figure}

Finally, we show the semimajor axis and eccentricity of each particle at the end of the simulation in Fig.\ \ref{fig:orbElem}. The solid vertical lines indicate the distance at which Phobos undergoes tidal disruption, which we define as the point at which its mass drops below half of its initial value. In addition, we show the analytic predictions for the Roche limit for the ellipsoid cases. The mass shedding limit (i.e., Eq. \ref{eq:d_shed}) is indicated by the dashed black line and the Drucker-Prager disruption distances are indicated with their respective dotted lines (i.e., Eq.\ \ref{eq:drucker}). Looking first at the ellipsoid case (Fig.\ \ref{fig:orbElem}a), we see that Phobos always disrupts well beyond the predicted distance based on the Drucker-Prager yield criterion. We attribute this discrepancy to two primary factors. {First, analytic estimates of the Roche limit do not capture the progressive weakening of the body as it migrates inward, whereas this effect is naturally incorporated into our discrete element approach through the dependence of bulk strength on the internal network of particle contacts. However, we note that the significance of this process depends on the size and shape distribution of constituent blocks, which are only narrowly explored here.} The second is that Eq.\ \ref{eq:C}, which is used to relate the inter-cohesive constant (which is an input parameter of the code) to the body's bulk cohesive strength, is only an approximation, which makes it difficult to make idealized comparisons between numerical and analytic estimates. The rubble piles used in this study likely have slightly lower bulk strengths than those estimated with Eq.\ \ref{eq:C}, as the bulk strength will vary throughout the body due to the random packing of particles. However, these simulations broadly agree with the analytic estimates, in the sense that the tidal disruption distance is relatively independent of the bulk cohesive strength at the Pascal level. {At kilopascal-level strengths, where cohesion is comparable to interior overburden pressures (tens of kilopascals), it becomes an important factor in setting the disruption distance.} In all cases, Phobos' tidal disruption would lead to the formation of a broad ring of particles having eccentricities ranging between ${\sim}10^{-2}$ and ${\sim}10^{-3}$.

\section{Discussion}\label{sec:discussion}
Derivations of the Roche limit that rely on static mechanics {can} underestimate the Roche limit{, depending on the scenario. In cases in which a body's cohesive strength is small compared to internal stresses from self-gravity, as may be the case for a rubble-pile Phobos, analytic treatments may be limited in their ability to account for gradual changes in surface and interior structure during inward migration. These effects, which may influence the bulk strength, are implicitly included in the numerical approach presented here.} The analytic treatments of \cite{Black2015}, \cite{Holsapple2008}, and others are useful for deriving insight or scaling relationships, but are limited to idealized ellipsoids and cannot account for more irregular shapes, which is shown to be an important effect here. In addition, an analytic treatment of Phobos Roche limit cannot account for time-varying stresses owing to Phobos' eccentricity and rotation. Today, Phobos has an eccentricity of ${\sim}0.015$ and a forced libration amplitude of ${\sim}1^\circ$ \citep{JacobsonR2010,Willner2010,Willner2014}.

\subsection{The ultimate fate of Phobos}
The approach considered here is effectively a best-case scenario in which Phobos is on a circular orbit in synchronous rotation. As Phobos migrates inward, it may encounter a number of semi-secular resonances with the Sun and/or tesseral resonances with the rotation of Mars that may excite its eccentricity \citep{Yokoyama2002,Yokoyama2005,Cuk2025PSJ}. If Phobos has substantial eccentricity (and therefore forced libration), time-varying tidal and rotational forces will make it even easier to strip mass from the sub- and anti-Mars points, possibly leading to more distant tidal stripping. We speculate that the excitation of Phobos' eccentricity at one of these resonances may play a significant role in determining its ultimate tidal disruption distance.

In any case, if Phobos can indeed shed mass prior to fully disrupting, then this material would enter circumplanetary orbit and re-impact Phobos at a later time. Owing to Phobos' excited orbit and differential nodal and apsidal precession between Phobos and any debris, these collisions would occur at speeds significantly higher than Phobos' surface escape speed, leading to runaway collisional erosion or “sesquinary catastrophe” \citep{Cuk2023}. Phobos will be particularly sensitive to this process as it migrates inward because its orbital velocity will increase, driving up the impact speeds. In addition, the increased tidal forces from Mars will reduce Phobos' strength against catastrophic collisions \citep{Agrusa2025b}. In such a scenario, the ultimate fate of Phobos may be more of a collisional process with the impactors sourced from the surface of Phobos itself. This outcome did not occur in our simulations because Phobos was forced onto a circular orbit and its tidal migration was artificially fast. In a more realistic scenario, in which Phobos has some eccentricity (and/or inclination) and migrates much more slowly, it may succumb to collisional bombardment before ever reaching the Roche limit. 

\subsection{Comparison with \cite{Black2015}}
\cite{Black2015} derived the disruption distance of Phobos as a function of its cohesion and arrived at results that are broadly consistent with those presented here. They used the same Drucker-Prager failure criterion for an ellipsoidal satellite derived by \cite{Holsapple2008} but concluded that Phobos will likely disrupt in the vicinity of ${\sim}1.6R_\M$. They arrived at this conclusion by assuming that Phobos cohesion is on the order of ${\sim}0.1 $ MPa, which comes from two lines of reasoning. First, the strength required to support topography on Phobos implies of a cohesive strength of at least ${\sim}0.01 $ MPa. Second, models of the Tagish Lake meteoroid fireball estimate a compressive strength of ${\sim}0.25$ MPa \citep{Brown2002}, which implies a global cohesive strength for Phobos of ${\sim}0.08$ MPa according to a Hoek-Brown material model. 

We argue that these strengths may be significantly overestimated. They estimated the minimum cohesive strength required to support topography as $P=\rho_\P g_\P h$, where $\rho_\P$ and $g_\P$ are Phobos' respective bulk density and surface gravity and $h$ is the maximum variation in surface height. This is an approximation commonly used for the surfaces of planets where the surface gravity is {approximately} constant. However, the surface gravity of Phobos varies by a factor of two \citep{Ernst2023a} over its surface and its bulk density may not be constant throughout the body, so this may not be a reliable lower limit for Phobos bulk cohesion. {The $0.01$~MPa cohesive strength required to support topography suggested by \cite{Black2015} may not be necessary according to the simulations presented here. In our rubble-pile models, with Phobos placed in orbit around Mars at Phobos' current distance, we do not observe noticeable changes to the global shape. Even in cohesionless cases, granular friction alone appears sufficient to preserve the body's overall structure against gravitational and rotational stresses. We note, however, that the limited particle resolution of this study precludes the resolution of finer-scale geographic features. Nevertheless, based on the highest-available-resolution shape model, gravitational slopes on Phobos generally do not exceed ${\sim}40^\circ$ \citep{Ernst2023a}, consistent with stability provided by friction without needing to invoke cohesion. At the same time, however, models proposed to explain Phobos' grooves typically invoke a weak, highly deformable interior overlaid by a regolith layer with a cohesion of tens of kilopascals \citep{Hurford2016,ChengBin2022}. Even our cohesionless simulations are compatible with such a scenario, as these groove-forming models invoke a cohesive surface layer having a thickness of ${\sim}100$ m, which is smaller than a single particle diameter in our simulations.
}

In addition, we argue that the strength of the Tagish Lake meteoroid does not provide a useful constraint for Phobos' cohesive strength for several reasons. First, the Tagish Lake meteoroid is only representative of Phobos if Phobos and Deimos are indeed primitive, carbonaceous objects captured around Mars, which is difficult to reconcile with their prograde and coplanar orbits \citep{Murchie2015}. Second, there is some evidence that Phobos and Deimos are similar to basaltic material in the mid-IR, which further complicates a capture origin \citep[e.g.][]{Glotch2018,Poggiali2022,Edwards2023}. Even assuming that the Tagish Lake meteorite is indeed a good representative, this does not necessarily constrain Phobos' bulk cohesive strength. For example, Tagish Lake also has some similarities with both Bennu and Ryugu \citep[e.g.,][]{Potiszil2024}, which were recently sampled by the OSIRIS-REx and Hayabusa2 missions. However, both of their surfaces are consistent with having little to no cohesive strength \citep{Arakawa2020,Walsh2022,Jutzi2022,Perry2022}. Similarly, the best meteorite analogs of the Didymos-Dimorphos system are L and LL chondrites \citep{deLeon2006,Dunn2013}, which have compressive strengths on the order of hundreds of megapascals \citep{Pohl2020}. Yet, the geology of the Didymos system and models of the DART impact are consistent with both bodies having cohesive strengths no higher than a few Pascals \citep[e.g.,][]{Daly2023,Raducan2024a,Barnouin2024a,Robin2024a,Bigot2024,Cheng2024}. Although these asteroids are very different than Phobos, this demonstrates that the strengths of individual components of a rubble-pile object do not necessarily correlate to a body's bulk cohesive strength. Rather, the body's bulk cohesive strength likely depends on the properties of a small-scale regolith (or lack thereof) that provides the “glue to hold all the stronger components together” \citep{Sanchez2014}. 

\subsection{Comparison with \cite{Madeira2023b}}
\cite{Madeira2023b} modeled the tidal disruption of Phobos using \textsc{pkdgrav}, the same code used in this study. However these simulations are sufficiently different to make it challenging to directly compare our results. They employed \textsc{pkdgrav} to verify that Phobos Roche limit matches their analytical estimate based on the Drucker-Prager yield criterion of an ellipsoidal satellite \citep{Holsapple2008}. Numerically, they find that a spherical Phobos with a friction angle of $\phi{\sim}40^\circ$ and a cohesive strength of 10 kPa would disrupt at a distance of 1.74 $R_\text{M}$. This result seems plausible, but it is difficult to compare directly to the results presented here, as we considered different values for the friction angle, cohesive strength, and bulk shape of Phobos. More importantly, these simulations instantaneously place Phobos on an orbit with a semimajor axis of $1.76R_\text{M}$ with its present-day eccentricity and Phobos disrupts on its very first orbit. They do not present a simulation in which Phobos does not disrupt, so these simulations do not seem to derive a limiting disruption distance. 

\subsection{The cyclic ring-satellite hypothesis}
It has been proposed that Phobos is an $N$-th generation product of an ongoing satellite-ring cycle \citep{Hesselbrock2017}. In this scenario, a satellite migrates inward due to tides until it disrupts at the rigid Roche limit (RRL) and forms a ring. Collisions of ring particles create an effective viscosity, which leads to the ring spreading as mass and angular momentum is transported throughout the disk \citep[e.g.,][]{Salmon2010}. This allows the ring to spread, where most material will land on Mars but some can spread beyond the fluid Roche limit (FRL), and reaccrete into a second-generation moon. Based on the density of Phobos, the FRL lies at approximately ${\sim}3.1R_\M$, while the synchronous orbit lies at ${\sim}6R_\M$. Although the disk torques can expand the orbit of newly accreted moonlets, they are not sufficient to reach the synchronous radius. This means that this process is cyclic; the moon's orbit will decay until it eventually disrupts and forms a new ring from which the next generation of satellite(s) will form. \cite{Cuk2020} demonstrate that Deimos' present-day low eccentricity and ${\sim}2^\circ$ inclination are consistent with an ongoing ring-moon cycle. If an ancient moon with ${\sim}20M_\text{Phobos}$ formed at the FRL and migrated outward due to torques from a ring, it would be captured into a 3:1 mean-motion resonance (MMR) with Deimos, exciting the latter's inclination to its present-day values. Because the ancient satellite would eventually reverse direction and migrate back inward, it would avoid crossing the 2:1 MMR that would have over-excited Deimos' eccentricity.

\cite{Madeira2023b} explored this possible ring-moon hypothesis and concluded that the ring-moon recycling process is dynamically viable. However, they argue that this process could not have occurred for Phobos, because a detectable ring system would be expected to persist today. Both the proposed model of \cite{Hesselbrock2017} and the rebuttal of \cite{Madeira2023b} rely on an assumed tidal disruption distance of each satellite generation at ${\sim}1.6-1.8R_\text{M}$. The work presented here demonstrates that a weak, Phobos-like satellite is more likely to disrupt in the vicinity of ${\sim}2-2.2R_\text{M}$, depending on its shape. A larger disruption distance would decrease the timescale of each satellite generation, and more importantly it would drastically increase the mass retained in each subsequent generation. {Under the assumption that Phobos (and earlier satellites) accreted from a Martian ring and have a weak, rubble-pile nature, it may be worth reevaluating models of an ongoing ring–moon cycle around Mars using a more realistic disruption distance.}

\subsection{Limitations of this study}
One key finding from these simulations is that the particle contact network evolves as Phobos migrates inward, leading to a reduction in cohesive strength. Due to computational constraints, however, the simulations use a narrow particle SFD and spherical particles, which do not capture interlocking effects. Future studies should aim to quantify how such effects could influence Phobos' tidal disruption limit. 

Additionally, the simulations assume a relatively homogeneous internal structure and material properties. Phobos may, in fact, possess a highly heterogeneous interior, with large contrasts in material properties and significant void spaces. Despite these uncertainties, this study provides a useful starting point for understanding Phobos' structural response to tidal forces. The upcoming JAXA MMX mission is expected to significantly refine knowledge of Phobos' properties, including surface regolith characteristics, mass distribution, and boulder SFDs \citep{Kuramoto2022}, enabling higher-fidelity models of its interior structure and tidal disruption.

\section{Conclusions}\label{sec:conclusions}

Using simple analytic estimates and numerical simulations, we have demonstrated that weak rubble piles can undergo tidal stripping well before they undergo complete tidal disruption. This is a result of tidal and rotational forces overcoming self-gravity at the sub-primary point before internal stresses trigger failure. Therefore, in cases in which cohesive forces are weak {relative to other stresses}, tidal disruptions may be governed by tidal stripping rather than internal failure. Once material is stripped from the surface, it enters planetocentric orbit, where it may collide with the satellite at a later time. Depending on the satellite's orbital excitation (eccentricity and inclination), this may lead to a runaway collisional erosion process that could destroy the satellite before the Roche limit is reached \citep{Cuk2023}.

We predict that this will be the case for Phobos, as it would be expected to be relatively weak if it reaccumulated from a disk of debris. When Phobos reaches ${\sim}2.2\RM$, material should be freely stripped from its surface. This could lead to sesquinary catastrophe if Phobos' eccentricity and/or inclination is sufficiently excited. If not, Phobos will still disrupt outside of ${\sim}2\RM$ unless it has a global minimum cohesive strength of 10s of kPa. Ignoring the possibility of impact-driven erosion, the exact disruption distance of Phobos will be highly sensitive to its interior structure and cohesive strength, which will be constrained by the MMX mission and its IDEFIX rover \citep[e.g.,][]{Kuramoto2022,Ulamec2025,Murdoch2025}.

\begin{acknowledgements}
    We thank Naomi Murdoch and Matija Ćuk for useful discussions, and the anonymous referee for their suggestions that significantly improved this manuscript. This work was supported by the French government through the France 2030 investment plan managed by the National Research Agency (ANR), as part of the Initiative of Excellence Université Côte d'Azur under reference number ANR-15-IDEX-01. The authors acknowledge support from CNES and The University of Tokyo and are grateful to the Université Côte d'Azur's Center for High-Performance Computing (OPAL infrastructure) for providing resources and support. Some simulations were also performed on the ASTRA cluster administered by the Center for Theory and Computation, part of the Department of Astronomy at the University of Maryland.
\end{acknowledgements}

\bibliographystyle{aa} 
\bibliography{references} 
\end{document}